\documentclass[9pt,journal]{IEEEtran}
\usepackage{xcolor}
\usepackage[english]{babel}

\usepackage{booktabs}       % professional-quality tables
\usepackage{makecell}
\usepackage{verbatim}
\usepackage{graphicx}
\usepackage{tabularx}
\usepackage{epsfig}
\usepackage{mathtools}
\usepackage{amsmath,amsthm,amsfonts,latexsym}    
\usepackage{cite}
\usepackage[switch]{lineno}
\usepackage[named]{algo}
\usepackage[noend]{algpseudocode}
\usepackage{algorithm}

\usepackage{balance}

\ifCLASSOPTIONcompsoc
  \usepackage[caption=false,font=normalsize,labelfont=sf,textfont=sf]{subfig}
\else
  \usepackage[caption=false,font=footnotesize]{subfig}
\fi

\usepackage{siunitx}

\usepackage[flushleft]{threeparttable} 

\newcommand{\abs}[1]{\lvert #1 \rvert}
\newcommand{\norm}[1]{\lVert #1 \rVert}

\newcommand{\minus}{\scalebox{0.75}[1.0]{$-$}}

\newtheorem{theorem}{Theorem}
\newtheorem{lemma}{Lemma}

\def\rr{{\mathbf r}}

\def\x{{\mathbf x}}

\def\e{{\mathbf e}}

\def\ee{{\boldsymbol\eta}}
\def\tt{{\boldsymbol\theta}}
\def\gg{{\boldsymbol\gamma}}

\def\DD{{\boldsymbol\Delta}}

\DeclareMathOperator*{\argmin}{\mbox{argmin}}
\DeclareMathOperator*{\argmax}{\mbox{argmax}}
\DeclareUnicodeCharacter{2212}{-}

\renewenvironment{IEEEbiography}[1]
  {\IEEEbiographynophoto{#1}}
  {\endIEEEbiographynophoto}
  
\begin{document}

\title{Constant Curvature Curve Tube Codes for Low-Latency Analog Error Correction}% for URLLC in 6G}

\author{Anders M. Buvarp~\IEEEmembership{Student Member,~IEEE},
        Robert M. Taylor Jr.~\IEEEmembership{Member,~IEEE}, 
        Kumar Vijay Mishra~\IEEEmembership{Senior Member,~IEEE},
        Lamine Mili~\IEEEmembership{Fellow,~IEEE} and
        Amir I. Zaghloul~\IEEEmembership{Life Fellow,~IEEE}
        \thanks{A. M. B., R. M. T. J., L. M., and A. I. Z. are with Bradley Department of Electrical and Computer Engineering, Virginia Tech, Blacksburg, VA 24061 USA. Email: rtaylor.rt@gmail.com,\{abuv,lmili,amirz\}@vt.edu.} 
        \thanks{K. V. M. and A. I. Z. are with the United States CCDC Army Research Laboratory, Adelphi, MD 20783 USA, e-mail: kvm@ieee.org, amir.i.zaghloul.civ@army.mil.}
        \thanks{K. V. M. acknowledges support from the National Academies of Sciences, Engineering, and Medicine via Army Research Laboratory Harry Diamond Distinguished Fellowship.}  
}

\maketitle

\begin{abstract}
    Recent research in ultra-reliable and low latency communications (URLLC) for future wireless systems has spurred interest in short block-length codes. In this context, we analyze arbitrary harmonic bandwidth (BW) expansions for a class of high-dimension constant curvature curve codes for analog error correction of independent continuous-alphabet uniform sources. In particular, we employ the circumradius function from knot theory to prescribe insulating tubes about the centerline of constant curvature curves. We then use tube packing density within a hypersphere to optimize the curve parameters. The resulting constant curvature curve tube (C3T) codes possess the smallest possible latency, i.e., block-length is unity under BW expansion mapping. Further, the codes perform within $5$ dB signal-to-distortion ratio of the optimal performance theoretically achievable at a signal-to-noise ratio (\textrm{SNR}) $< \minus5$ dB for BW expansion factor $n \leq 10$. Furthermore, we propose a neural-network-based method to decode C3T codes. We show that, at low \textrm{SNR}, the neural-network-based C3T decoder outperforms the maximum likelihood and minimum mean-squared error decoders for all $n$. The best possible digital codes require two to three orders of magnitude higher latency compared to C3T codes, thereby demonstrating the latter's utility for URLLC.
\end{abstract} 

\section{Introduction}
\label{sec:intro}
Digital communication systems have relied upon Shannon’s separation principle \cite{Shannon1948,Cover1991} to separate source and channel coding allowing bits to act as the ``universal currency''. However, using bits as the representation of data for transmission and storage is known to be sub-optimal when the source and channel codes are finite length or the channel is non-ergodic (see, e.g., \cite{GastparMay2003,Gastpar2006}). Many current and evolving real-time applications such as autonomous driving, tactile internet, telerobotics, audio-visual conferencing, and industrial robotic control require minimal latency. Therefore, large block-lengths typical in digital capacity-approaching codes are not desirable in these applications. The fifth-generation New Radio (5G-NR) cellular standards committee recognizes this problem in regard to the emerging ultra-reliable low latency communications (URLLC) and there is significant interest in short block-length digital channel codes (see, e.g., \cite{Shirvanimoghaddam2019}). 

However, sometimes, it is more desirable to build a communications system that handles continuous signals  without source quantization. Among other benefits (see, e.g., \cite{ChenWornell1998} for further details), analog coding strategies may allow \textit{graceful degradation} at the receiver, a useful characteristic for broadcast systems. Previous research \cite{Huang2018,Zhang2020} mentions ultra-reliable finite length codes that achieve high throughput and use arbitrary code-rate analog fountain codes with linear computational complexity. Similarly, another study in \cite{Shirvan2013} proposed capacity-approaching analog fountain codes. In this paper, we focus on mapping analog source samples directly to channel inputs and hence achieve minimal latency (arising from unity block-length) while still retaining high source reconstruction accuracy. 

We apply the general framework of joint source-channel coding to independent and identically distributed (i.i.d.) continuous alphabet uniform sources and focus solely on analog error-correcting codes (thus neglecting the analog source coding component). Here, analog error correction refers to a type of Shannon-Kotel\'{n}ikov (SK) joint source-channel code \cite{Shannon1949,Kotelnikov1959} applied to continuous alphabet sources, where the channel bandwidth (BW) exceeds the source signal BW. We consider integer-valued BW expansion maps whereby a given memoryless uniform  source symbol is mapped onto an $n$-dimensional vector of continuous  values for transmission (or storage) over an i.i.d. Gaussian channel.  

The BW expansion SK maps have been studied in detail \cite{Hekland_thesis2007,Hekland2009,Akyol2010,Saleh2011,Campello2012,Campello2013}, wherein the curves were generally variations of a 2-D Archimedes spiral. Other approaches include chaotic dynamics \cite{ChenWornell1998}, orthogonal polynomials \cite{VaishampayanCosta2003}, and \emph{geodesics} on flat tori \cite{Kravtsov2007,Wernersson2009,Xie2010}. Here, recall that geodesic is a curve that represents the shortest arc between two points on a \emph{Riemannian manifold} --- a real differentiable manifold in which the tangent space at each point is a finite-dimensional Hilbert space and, hence, provided with an inner product. However, a general SK map theory for high-dimension expansions remains relatively unexamined. In \cite{Saleh2011}, a $1$:$3$ BW expansion approach combined a scalar quantizer with a 2-D Archimedes spiral. But the gap to the optimal performance theoretically achievable (OPTA) \cite{Berger1967} was significantly large.  Analog channel coding has been successfully employed for low signal-to-noise ratio \textrm{SNR} (especially near $0$ dB) using Gaussian sources, $1$:$2$ BW expansion with non-linear curves, and MMSE decoding \cite{Hu2011}.  Low-delay joint source-channel coding with autoencoders has been shown to operate well for low \textrm{SNR} using Gaussian sources \cite{Xuan2023}. 

A theoretical framework for a high-dimension SK map was first proposed in \cite{Hekland2009}. They proved that $1$:$n$ analog codes formed by densely packing tubes inside a hypersphere (for average power constraint) led to performance gains of $n$ dB source-to-distortion ratio (\textrm{SDR}) for each $1$ dB of channel \textrm{SNR} for uniformly distributed sources in additive white Gaussian noise (AWGN) channels under high \textrm{SNR} regime. This performance slope follows the slope of OPTA curves and is therefore worthy of consideration. 

Our previous work \cite{Taylor2013} showed that optimal BW expansion encoding maps vary as a function of \textrm{SNR} and, therefore, code selection is a function of the target channel \textrm{SNR}. An insulation property of curves (sometimes, referred to as the \textit{small-ball radius}) is necessary to prevent self-intersections or tight bending that may lead to the threshold effect \cite{Shannon1949,Hekland2009} in analog coding/modulation. Thus, code design for $1$:$n$ maps entails constructing 1-D manifolds (curves) embedded in a $n$-D Euclidean space in such a  way that the curve  maintains a certain distance from itself while fitting inside or on a $n\num{-1}$ sphere. This type of code design is essentially \textit{tube packing} \cite{Taylor2013}, where the centerline of the tube is the locus of our analog encoder map. Our previous work \cite{Taylor2013} considered only the case of \emph{two} degrees of freedom with $n=4$ and excluded \textrm{SNR} below $1$ dB. Previously, in \cite{taylor2017structured}, we considered spherical codes with only odd dimensions. In this paper, we develop codes for arbitrary $n$, set the associated frequency parameters to harmonic integer multiples, optimize the curve parameters using the simultaneous perturbation stochastic approximation (SPSA) algorithm, evaluate the performance for very low \textrm{SNR} levels down to $\minus10$ dB, implement a neural-network-based decoder, and include a comparison with digital codes.

In this work, we leverage upon the tube packing \cite{Taylor2013} and knot theory \cite{Gonzalez1999} to design $1$:$n$ analog  codes for arbitrarily high $n$-dimensional spaces. Contrary to prior works on analog error correction that focused primarily on the high \textrm{SNR} regime and Gaussian sources \cite{VaishampayanCosta2003}, our work is applicable to very low \textrm{SNRs} and considers uniformly distributed sources. In this work, our main contributions are:\\
\textbf{1) Minimum latency.} We design constant curvature curve tube (C3T) codes for analog error correction that comes within $5$ dB \textrm{SNR} of OPTA for $n \leq 10$ and \textrm{SNR} $<\minus5$ dB. The codes require only one source sample for each code block (minimum possible encoding latency). Similar to all constant curvature curve codes studied in the past, our codes are also orthogonal polynomials by virtue of the harmonic frequency structure selected \cite{Kravtsov2007}. Some parametric curves such as Archimedes' spiral mapping \cite{Hekland2009} may similarly achieve minimal latency. However, C3T codes also perform better at low SNR. \\
\textbf{2) Tube packing.} We use the global radius of curvature \cite{taylor2017structured} formulation as our objective function to pack tubes on a flat torus for optimizing the parameters of a constant curvature curve encoder map. We use simultaneous perturbation stochastic approximation as our optimizer because analytic gradients do not exist.\\
\textbf{3) Validation through neural networks.} We train and deploy a hardware-accelerated multi-layer perceptron (MLP) network to provide precise closed-form neural network decoders as an approximation for our proposed analog codes. We also develop the Bayesian MMSE and maximum \textit{a posteriori} (MAP) decoders to compare the performance of our C3T codes with Archimedes' spiral codes\cite{Hekland2009}, hybrid scalar quantizer linear coder (HSQLC) \cite{Kleiner2009}, hybrid analog-digital codes \cite{Saleh2011} and spherical codes \cite{VaishampayanCosta2003}. \\
\textbf{4) Source sample reduction.} We demonstrate significant improvements over the theoretically optimal digital channel coding scheme in terms of the required number of source samples needed to achieve a (\textrm{SNR}, \textrm{SDR}) pair.

Ideally, communications should be instantaneous and without errors. The URLLC service area within 5G-NR requires packet delays of less than $1$ millisecond with extremely high reliability, $99.999\%$ [6]. Our C3T codes do not require source coding, packetization, and buffering thereby easily meeting the low latency requirement. Further, the minimum MMSE decoder for our C3T codes yields very high ($99.96\%$) accuracy. Compared to highly complex digital systems that are in use nowadays, C3T codes are simpler to design. 

The rest of this paper is organized as follows. In the next section, we provide the mathematical preliminaries for constant curvature codes. Section~\ref{sec:constant_curvature_curves} introduces our C3T codes and provides the decoder in Section~\ref{sec:decoder}. We validate the performance of C3T codes using digital vs analog coding block-length comparisons through numerical experiments in Section~\ref{sec:numexp}. We conclude in Section~\ref{sec:summ}.

Throughout this paper, we use the following notations: $\mathbb{R}$ is the set of real numbers, $\mathbb{Z}$ is the set of integers, $\mathbb{N}^0$ is the set of whole numbers, $\mathbb{N}^+$ is the set of natural numbers, $\mathbb{E}^{n}$ is the $n$-dimensional Euclidean space, $\mathrm{E}$ is the expectation function, $\textbf{C}_{yy}$ is the covariance matrix, $\textbf{C}_{sy}$ is the cross-covariance matrix, $\mathcal{O}$ is the Bachmann-Landau big $O$ notation, $C^r$ is an $r$ times continuously differentiable parametric curve, $\mathbb{T}^{n/2}$ is a $n/2$-dimensional flat torus, $\mathcal{S}^{n/2}$ is a $n/2$-dimensional unit sphere, $\langle \cdot,\cdot \rangle$ denotes the inner product, $\angle(\cdot,\cdot)$ denotes the angle between two vectors, $\lVert \cdot \rVert_1$ is the $\ell_1$-norm, $\lVert \cdot \rVert$ is the $\ell_2$-norm, $P_{(\cdot)}$ denotes projection matrix, $P^\perp_{(\cdot)}$ denotes orthogonal (or perpendicular) complement, $P^T$ denotes the matrix transpose, $\mathbf{1}_{n}$ is a vector of all ones of size $n$, $\mathbf{0}_{n}$ is a vector of all zeros of size $n$, $\textbf{y}$ is a column vector, $\overline{y}$ is the mean of $\textbf{y}$, $\mathbf{I}$ is the identity matrix, $((\cdot))_q$ is the modulo-q operator, $\lceil \cdot \rceil _q$ denotes the ceiling function, diag$(\dots)$ is the diagonal matrix, and $x^{(k)}(\alpha)$ is the $k^{(th)}$ derivative of $x$ with respect to $\alpha$. The notation $x \sim \mathcal{U}(a,b)$ means a random variable drawn from the uniform distribution over the interval $[a,b]$ and $x \sim \mathcal{N}(\mu,\sigma^2)$ represents the Gaussian distribution with mean $\mu$ and variance $\sigma^2$. A random variable drawn from the Bernoulli distribution is represented by $x \sim \mathcal{B}(p)$.

\section{Desiderata for Constant Curvature Codes}
\label{sec:prelim}

A $n$-dimensional manifold is a Hausdorff and second countable topological space which is locally homeomorphic to Euclidean space of dimension $n$. A curve $\mathbf{x}$ is a vector-valued bijective function that maps an
interval $I$ of the real line to a 1-D manifold $\mathcal{M}_1$ embedded in $n$-dimensional Euclidean space such that
\begin{align}
  \mathbf{x} : I \rightarrow \mathcal{M}_1 \subset \mathbb{R}^n,
\end{align}
where $\mathbf{x}$ is a parametric curve of class $C^r$. The curve $\mathbf{x}$ is parameterized by $\alpha \in I$ and $\mathbf{x}(I)$ is the image of the curve. A $C^r$ curve is regular of order $m$ if, for any $\alpha \in I$, \{$\mathbf{x}^\prime(\alpha)$, $\mathbf{x}^{\prime\prime}(\alpha)$, \dots, $\mathbf{x}^{(m)}(\alpha)$\}, $m \leq k$ are linearly independent vectors in $\mathbb{R}^n$. We define the Frenet frame for the regular curve $\mathbf{x}(\alpha)$ as the set of orthonormal vectors $\mathbf{e}_1(\alpha)$, \dots, $\mathbf{e}_n(\alpha)$ formed as
\begin{align*}
    \mathbf{e}_1(\alpha) = \mathbf{x}^\prime(\alpha) / \lVert \mathbf{x}^\prime(\alpha) \rVert, \quad \mathbf{e}_k(\alpha) = \frac{\tilde{\mathbf{e}}(\alpha)}{\lVert \tilde{\mathbf{e}} (\alpha) \rVert},
\end{align*}
\begin{align}
    \tilde{\mathbf{e}}_k(\alpha) = \mathbf{x}^{(k)}(\alpha) - \sum_{i=1}^{k-1} \Big < \mathbf{x}^{(k)}(\alpha),\mathbf{e}_i(\alpha) \Big > \mathbf{e}_i(\alpha),
    \label{equ:frenet_frame}
\end{align}
from Gram-Schmidt orthogonalization. The \emph{generalized curvatures} are then $n\num{-1}$ real-valued functions % $\chi_m(\alpha)$ given by
\begin{align}
    \chi_m(\alpha) = \frac{\langle \mathbf{e}^\prime_m(\alpha),\mathbf{e}_{m+1}(\alpha) \rangle}{\lVert \mathbf{x}^\prime (\alpha) \rVert}, \quad m = 1, 2, \dots, n\num{-1} \quad .
    \label{equ:generalized_curvatures}
\end{align}

Define $\mathbf{B}$ a skew-symmetric matrix composed of the generalized curvatures as
\begin{align}
    \mathbf{B} = 
    \begin{bmatrix}
    0 & \chi_1(\alpha) & \dots & 0 & 0 \\
    -\chi_1(\alpha) & 0 & \dots & 0 & 0 \\
    \vdots & \vdots & \ddots & \vdots & \vdots \\
    0 & 0 & \dots & 0 & \chi_{n\num{-1}}(\alpha) \\
    0 & 0 & \dots & -\chi_{n\num{-1}}(\alpha) & 0
    \end{bmatrix}.
\end{align}
From (\ref{equ:generalized_curvatures}), we then obtain a system of first-order differential equations, known as the Frenet-Serret formula, as
\begin{align}
    \lbrack \mathbf{e}^\prime_1(\alpha),\dots, \mathbf{e}^\prime_n(\alpha) \rbrack ^T = \lVert \mathbf{x}^\prime (\alpha) \rVert \; \mathbf{B} \; \lbrack \mathbf{e}_1(\alpha),\dots,\mathbf{e}_n(\alpha)\rbrack ^T.
    \label{equ:frenet-serret}
\end{align}

The fundamental theorem of curves \cite[pp. 84-85]{banchoff2010differential} states that given $n\num{-1}$ parametric functions $\chi_m$ with ${\chi_m(\alpha)} > 0, m = 1, 2,\cdots, n\num{-1}$, there exists a regular $C^n$ curve $\mathbf{x}$ which is unique up to transformations under the Euclidean group. A Euclidean group is the group of Euclidean isometries of a Euclidean space $\mathbb{E}^{n}$; that is, the transformations of that space that preserve the Euclidean distance between any two points (also called \textit{Euclidean transformations}). This means we can translate, rotate, or scale the image of the curve, thereby modifying the parameters of the map accordingly, without changing the generalized curvatures.

\subsection{Flat tori} 
The \emph{Gaussian curvature} of a surface at a point is the product of two principal curvatures at the given point, where the two principal curvatures at a given point of a surface are the eigenvalues of the shape operator at the point.  A $n/2$-dimensional flat torus $\mathbb{T}^{n/2}$ is a closed Riemannian manifold that has zero Gaussian curvature everywhere. It is defined as the product of $n/2$ circles, and algebraically, as a quotient space $\mathbb{R}^n$ / $\mathbb{Z}^n$. For more details on a flat torus, we refer the reader to seminal works by Costa et. al. \cite{Costa2004,Costa2018} and references therein. 

Since $\mathbb{T}^{n/2}$ flat tori are sub-manifolds of the ($n − 1$)-sphere, packing tubes on a flat torus means the tubes are also being packed on a sphere. This ensures the channel input obeys maximum and average power constraints. The fact that constant curvature curves are geodesics on flat tori allows for improved decoder performance since we can project observations onto the torus prior to decoding to the closest point on the curve. That $\mathbb{T}^{n/2}$ is locally isometric \cite{Coxeter1969} with a $n/2$-hyperrectangle establishes a link to lattice coding methods, but those are not used here; more details on this are available in our previous work \cite{taylor2017structured}. Nearly all prior studies \cite{VaishampayanCosta2003,Campello2013} on analog coding with constant curvature curves explicitly make use of the local isometry with the hyperrectangle in their decoder construction.

Recall the following useful result on computing some geometric properties of the hypersphere that we use later to design C3T codes.
\begin{lemma}\cite{ConwaySloane1998}
    The volume $V_n$ of a $n$-sphere with radius $R$ is
    \begin{align}
        V_n(R) = C_n R^n,
        \label{equ:V_n(R)}
    \end{align}
    where
    \begin{align*}
      C_n = \frac{\pi^{n/2}}{\Gamma (\frac{n}{2}+1) } = 
      \begin{dcases}
        \frac{\pi^{n/2}}{(n/2)!}, &\text{n even,} \\
        \frac{1}{n!!}2^{\frac{n+1}{2}} \pi^{\frac{n\num{-1}}{2}}, &\text{n odd,}
      \end{dcases}
    \end{align*}
    and where $\Gamma (t) = \int_{0}^{\infty} x^{t-1} e^{-x} dx $ is the gamma function \cite{Gonzalez1999}. The surface area is then
    \begin{align}
        S_n = \frac{R^2n\pi^{n/2}}{\Gamma (\frac{n}{2}+1) } \quad. 
    \end{align}
\label{lemma:hypervol}
\end{lemma}

%-----------------------------------------------------------------------------------
\begin{figure}[t]
\centering
    \includegraphics[width=\columnwidth]{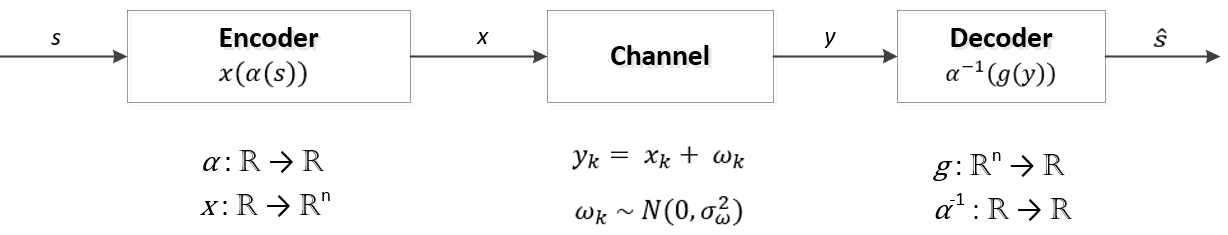}
    %\vspace{-4cm}
    \caption{Block diagram of joint source-channel analog coding architecture. A source $s$ is encoded to $x$, passed through a channel to create $y$, and decoded to recover the source estimate $\hat s$.}
    \label{fig:system_block_diagram}
\end{figure}
%-----------------------------------------------------------------------------------

\subsection{System model}
Fig.~\ref{fig:system_block_diagram} illustrates the three basic components of our analog error correction system. Consider AWGN channels such that $\mathbf{y} = \mathbf{x} + \mathbf{w}$ where $\mathbf{w} \sim \mathcal{N}(0,\sigma^2_w \mathbf{I})$, $s \sim \mathcal{U}(a,b)$, and $\mathrm{E}[x^2_m] = P$. The encoder is a curve composed of a scalar-valued \emph{stretching} function $\alpha (s)$ with a vector-valued \emph{folding} function $\mathbf{f}$ that together maps a continuous source symbol $s$ into the channel input $\mathbf{x}$ as $\mathbf{f}: \alpha(s) \rightarrow \mathbf{x}$ with an effective code block-length of unity. The decoder is a surjective map composed of an \emph{unfolding} function $g$ containing a vector input with an \emph{unstretching} function $\alpha^{-1}(\cdot)$ that together maps the channel output $y$ into source estimate $\hat \alpha$ as $g: \mathbf{y} \rightarrow \hat \alpha$.

We evaluate the performance of our analog coding scheme using OPTA \cite{Berger1967} to relate the source reconstruction mean-square error distortion $D$ to the input channel average power $P$ using the rate-distortion function \cite{Cover1991},
\begin{align}
    R(D) = nC(P),
\end{align}
where $C(P)$ is the capacity cost function \cite{GastparMay2003}. % and $R(D)$ is the rate-distortion function \cite{Cover1991}. 
For AWGN channel with noise variance $\sigma^2_w$ and average power $P$, we have $C(P) = 0.5\ln(1 +
P/\sigma^2_w)$. Since there is no closed-form expression for $R(D)$ given a uniform source with mean-square error distortion, we can approximate $R(D)$ with the following Shannon lower bound \cite{Berger1971} using the entropy $h(s)$ of the source $s$:
\begin{align}
    R(D) \geq h(s)-\frac{1}{2}\ln(2\pi eD).
\end{align}
For $s\sim \mathcal U[a,b]$, source variance is $\sigma^2_s = (b − a)^2/12$ and $h(s) = \ln(b−a)$; we refer the reader to \cite{Cover1991} for entropies of various source distributions. The \emph{upper bound} on \textrm{SDR}=$\sigma^2_s/D$ as a function of \textrm{SNR}=$P/\sigma^2_w$ is,
\begin{align}
    \textrm{SDR} \leq \frac{\pi e}{6}(1+\textrm{SNR})^n.
    \label{equ:opta}
\end{align}
For non-Gaussian sources \cite{Saleh2011}, the lower bound corresponds to the
$R(D) = 0.5\ln(\sigma^2_s / D)$ computed for Gaussian i.i.d. sources \cite{Cover1991}, i.e., 
 %and corresponds to the \emph{lower bound} on the OPTA for any non-Gaussian i.i.d. source \cite{Saleh2011}, that is,}
\begin{align}
    \textrm{SDR} \geq e^{2nC}.
    \label{equ:opta_lower_bound}
\end{align}

For digital coding schemes the only source of distortion on an i. i. d. uniform source stream is the quantization noise. For a uniformly distributed source, the source-to-quantization noise value in decibels for a ($B+1$)-bit quantizer of a source, $s \in \mathcal U$, with variance $\sigma^2_s$ is well known to be \textrm{SDR}$_{\textrm{dB}} = 10 \log_{10}(12\sigma^2_s 2^{2B} / S^2_m) \approx 6.02B$. Therefore, the average number of bits per source symbol is well approximated as \textrm{SDR}$_{\textrm{dB}}/6.02 + 1$.

For the code block-length $N_c$, the Polyanskiy bound \cite[Eq. (1)]{Polyanskiy2010} provides an expression for the source rate $R$ as a function of the codebook size $M$, capacity $C = 0.5 \log_2(1 + \textrm{SNR})$, block error rate $\epsilon$, and channel dispersion $V$ as
\begin{align}
\label{eq:polyanskiy}
    R = \log_2(M)/N_c \approx C - \sqrt{\frac{V}{N_c}} \: Q^{-1}(\epsilon),
\end{align}
where $Q(x) = 1/\sqrt{2\pi} \int_{x}^{\infty} e^{-u^2/2}du$ is the Gaussian right-tail probability. For an AWGN channel, the channel dispersion is \cite[Eq. (293)]{Polyanskiy2010}
\begin{align}
\label{eq:ch_disp}
    V = \frac{\textrm{SNR}}{2}\frac{(\textrm{SNR}+2)}{(\textrm{SNR}+1)^2}(\log_2e)^2.
\end{align}
Substituting \eqref{eq:ch_disp} in \eqref{eq:polyanskiy} and solving for $N_c$ yields the following expression for the minimal digital channel code block-length:
\begin{align}
\label{equ:pb}
    N_c = \frac{(\log_2e)^2}{2(C-R)^2}\left ( 1-\frac{1}{(1+\textrm{SNR})^2}\right )(Q^{-1}(\epsilon))^2.
\end{align}

\section{C3T Codes}
\label{sec:constant_curvature_curves}
The functional form of curves of constant curvature is computed by solving the Frenet-Serret equation (\ref{equ:frenet-serret}). Constant curvature curves in even dimensions are a class of curves that are geodesics on flat tori $T^{n/2}$ embedded in an $n\num{-1}$ sphere, which itself is embedded in an $n$-dimensional Euclidean space. These curves are represented as the vector \cite{Costa1990}
\begin{align}
\mathbf{x}_{\tt}(\alpha)&=[r_1\cos(\omega_1 \alpha), r_1\sin(\omega_1 \alpha),  
r_2 \cos(\omega_2 \alpha), r_2 \sin(\omega_2 \alpha), \nonumber \\ 
& ...,r_{n/2} \cos(\omega_{n/2} \alpha), r_{n/2} \sin(\omega_{n/2} \alpha) ]^T,
\label{equ:parametric_curve}
\end{align}
that is parameterized by the vector $\theta=[r_1,r_2,...,r_{n/2},\omega_1,\omega_2,...,\omega_{n/2}]^T$, where $r_i$ and $\omega_i$, $i=1,\cdots, n/2$, are radius and frequency parameters, respectively. Recall the following useful result from %Theorem~\ref{theorem:constant_curvature} follows a claim in 
\cite{Costa1990} that states that the geodesics on flat tori must all have constant generalized curvature.
%(stated therein without proof but found to be correct in our proof in Appendix~\ref{sec:constant_curvature_proof})  
\begin{theorem}\cite{Costa1990}
 The curve in (\ref{equ:parametric_curve}) is a constant curvature curve.
\label{theorem:constant_curvature}
\end{theorem}
%\begin{IEEEproof}
%\textcolor{red}{For any curve constituted as in \eqref{equ:parametric_curve}, given two arbitrary points, there is an isometry of $\mathbb{R}^3$ onto itself.  %This isometry takes the curve onto itself and the one point onto the other. 
%    Therefore, the curvature of the curve is the same at all points \cite{Gluck1966}. Additionally, \eqref{equ:parametric_curve} is also an embedding with orthogonal basis in an $n$-dimensional Euclidean space \cite{Costa2004}.  This is also applicable to the odd-dimensional case. }
%\end{IEEEproof}

Constant curvature curves in odd dimensions are generalized helices of a very similar form but with an additional real-valued scalar parameter $b$ as 
\begin{align}
\mathbf{x}_{\tt}(\alpha)&=[r_1\cos(\omega_1 \alpha), r_1\sin(\omega_1 \alpha),  
r_2 \cos(\omega_2 \alpha), r_2 \sin(\omega_2 \alpha), \nonumber \\ 
& ...,r_{(n\num{-1})/2} \cos(\omega_{(n\num{-1})/2} \alpha), r_{(n\num{-1})/2} \sin(\omega_{(n\num{-1})/2} \alpha), b\alpha ]^T,
\label{equ:parametric_curve2}
\end{align}
with parameter vector $\mathbf{\theta}=[r_1,r_2,...,r_{(n\num{-1})/2},\omega_1,\omega_2,...,\omega_{(n\num{-1})/2}, b]^T$.

To ensure our encoder map locus has a bounded channel input power, we enforce that the constant curvature curve must lie on or inside a unit $n\num{-1}$ sphere such that
\begin{align}
  \norm{\mathbf{x_{\theta}}(\alpha)} =
    \begin{dcases}
        \sum_{i=1}^{n/2}r_i^2 = 1, &\text{n even,} \\
        \sum_{i=1}^{(n\num{-1})/2}r_i^2 + \pi^2b^2 \leq 1 , &\text{n odd.}
    \end{dcases} 
\end{align}
When the radii are equal for all even $n$, we have
$\theta=[r_1,\cdots,r_{n/2}]^T=\sqrt{2/n}\mathbf{1}_{n/2}$.  We
compactly write the $k$-th derivative of the $m$-th component of
this curve for $m=1,2,\cdots,n$ as
\begin{align}
x_{m,\theta}^{(k)}(\alpha)=r_{\lceil \frac{m}{2} \rceil}\omega_{\lceil \frac{m}{2} \rceil}^{(k)}
  \cos\left(\omega_{\lceil \frac{m}{2} \rceil}\alpha+\frac{\pi}{2}\left (k-((m-1))_2 \right )\right) 
  \;\;\;.
\label{equ:curve_derivative}
\end{align}

It follows that, to achieve full dimensionality, $r_i>0$. Furthermore, for the curve to be ``twisted'' (meaning all curvatures are greater than zero), the frequency parameters $\omega_i$ must all be different from each other \cite{Costa1990}. By restricting the frequency parameters $\mathbf{\omega}=\{\omega_1,\cdots,\omega_{n/2}\}$ to be harmonic, that is integer multiples of the fundamental frequency $\omega_1$,
\begin{equation}
\omega_i=z_i\omega_1\;\;, z_i\in\mathbb{Z} \;\;\;i=2,3,\cdots,n/2,
\label{equ:harmonic}
\end{equation}
we obtain curves that repeat themselves. The curve is, thus, closed or, in other words, an $n$-dimensional \textit{knot} is formed.  We prevent the ending point of the curve from coming back to the starting point by limiting its path length to correspond to an amount less than the full path length. In the sequel, we compute tube densities for various choices of integer-multiple frequency parameters.

The path length $L$ for constant curvature curves is
\begin{align}
  L = \int_{\alpha=-\pi}^{\pi} \norm{\mathbf{x_{\theta}^\prime}(\alpha)} d\alpha = 
    \begin{dcases}
        2\pi \sqrt{\sum_{i=1}^{n/2} \omega_i^2 r_i^2}, &\text{n even,} \\
        2\pi \sqrt{\sum_{i=1}^{(n\num{-1})/2} \omega_i^2 r_i^2 + b^2}, &\text{n odd.} 
    \end{dcases}
\label{equ:L}
\end{align}

Constant curvature curves have constant stretch and the path length
from a given point on the locus is linear in $\alpha$. These curves will serve as the encoder function for analog error correction as illustrated in Fig. \ref{fig:system_block_diagram}. To finish specifying the encoder map, we compute the parameters $\mathbf{\theta}$ of the curves in (\ref{equ:parametric_curve}) and
(\ref{equ:parametric_curve2}).

\subsection{Knot-theoretic formulation}
\label{sec:knot}
The aforementioned theory of curves is not enough to establish our method of
tube packing because 1-D manifolds do not have any inherent thickness property. For tube packing, such an insulation property is necessary to prevent self-intersections or tight bending that leads to the
\emph{threshold effect} \cite{Shannon1949,Hekland2009}.  We desire a
mathematical tool that conveys the idea of insulation and prevents
our tubes from self-intersecting or bending too tightly.  The global
radius of curvature introduced in Gonzalez
et. al. \cite{Gonzalez1999} in the context of knot theory is one
such way to ascribe an insulating radius about the center line of a
curve.

%-----------------------------------------------------------------------------------
\begin{figure}[t]
\centering
  \subfloat[ ]{\includegraphics[width=0.45\columnwidth]{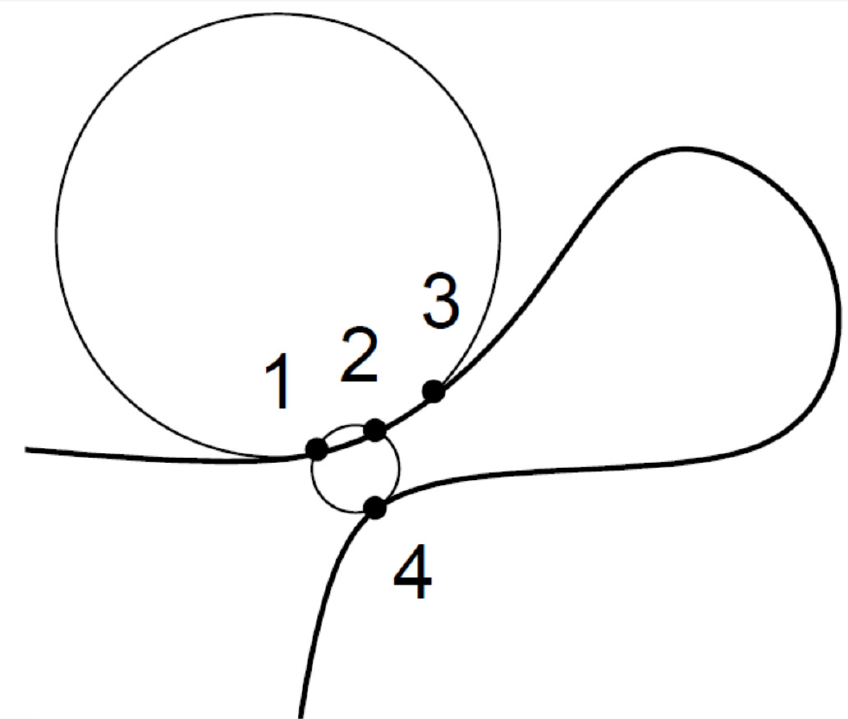}}
  \hfil
  \subfloat[ ]{\includegraphics[width=0.5\columnwidth]{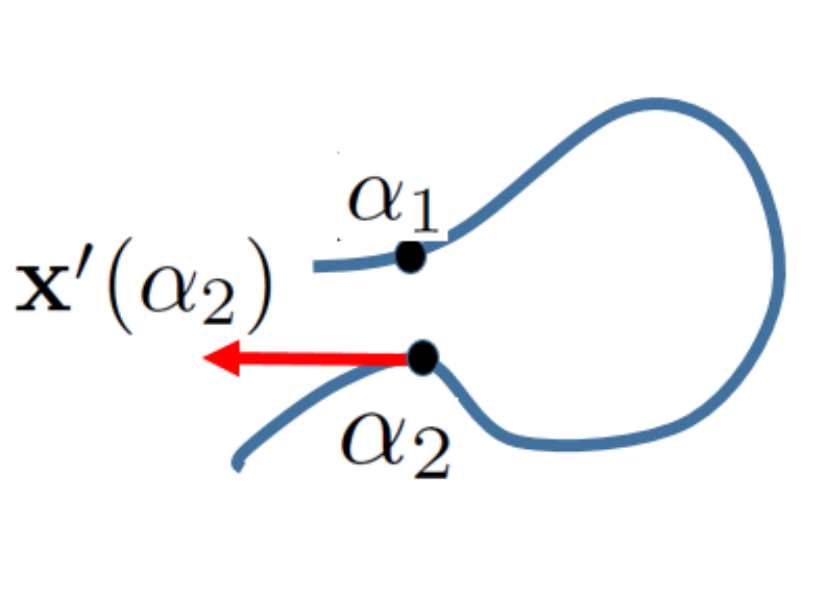}}
  \caption{(a) The radius of the circumcircle through points 1,2, and 3 approximates the local radius of curvature at point 2, whereas the global radius of curvature at point 2 contains information concerning non-local parts of the curve. This figure and caption are from \cite{Gonzalez2002}. (b) Two points on the curve and the tangent vector at one of those points are sufficient to compute the circumradius.}
    \label{fig:global_radius_of_curvature_pic}
\end{figure}
%-----------------------------------------------------------------------------------

Following \cite{Gonzalez2002}, Fig.~\ref{fig:global_radius_of_curvature_pic} illustrates circumcircle and its associated circumradius. The smallest circumradius taken over all possible triplets of points on the curve yields the global radius of curvature and consequently the tube radius. The circumradius of a curve using tangent-point form is
\begin{align}
\rho(\alpha_1,\alpha_2) =\frac{\norm{\mathbf{x}(\alpha_1)-\mathbf{x}(\alpha_2)}}
 {2\abs{\sin \left(\angle\left(\mathbf{x}(\alpha_1)-\mathbf{x}(\alpha_2),\mathbf{x}^\prime(\alpha_2)\right)\right)}}.
\label{equ:circumradius}
\end{align} 
We state the following useful lemmata that we use later to compute local curvature. Define the coefficient $c_{kjl}$, which is a constant for fixed encoder parameters and auxiliary variables $\lambda_{ji}$, such that
\begin{equation}
  c_{kjl}=\left( \sum_{i=1}^{n/2}\lambda_{ji}^lr_i^2\omega_i^{k+j} \right)^{1/2}.
\end{equation}
Using this notation, we write $\norm{x_{\theta}(\alpha)}=c_{000}$ and 
$\norm{x_{\theta}'(\alpha)}=c_{110}$.  More generally, we have
$\norm{x_{\theta}^{(k)}(\alpha)}=c_{kk0}$.
%Lemma~\ref{lemma:frenet_frame_curve_relation} 
\begin{lemma}
For a set of arbitrary real diagonal matrices
$\Lambda_j=\mbox{diag}(\lambda_{j,1},\lambda_{j,2},\cdots,\lambda_{j,n/2})$,
$j\in\mathbb{N}^0$, $k\in\mathbb{N}^+$, $p\in \mathbb{Z}$,
and for the curve given in (\ref{equ:parametric_curve}), the following
holds:
\begin{align}
\langle x_{\theta}^{(k)},\Lambda_j x_{\theta}^{(j)} \rangle
=\left\{  \begin{array}{cc}
0, & k-j=2p-1, \\
c_{kj1}^2, & k-j=4p, \\
 -c_{kj1}^2, & k-j=4p-2.
\end{array} \right.
\label{equ:inner_product_relation}
\end{align}
\label{lemma:inner_product_relation}
\end{lemma}
\begin{IEEEproof}
Using (\ref{equ:curve_derivative}) and simple trigonometric
identity $\cos(A-B)=\cos(A)\cos(B)+\sin(A)\sin(B)$ we write \begin{align}
 & \langle x_{\theta}^{(k)},\Lambda_j x_{\theta}^{(j)} \rangle = \nonumber \\
 &  \sum_{i=1}^{n/2} \lambda_{ji}r_i^2\omega_i^{k+j}\cos(\omega_i\alpha+
    \frac{\pi k}{2})\cos(\omega_i\alpha+\frac{\pi j}{2})+  \nonumber \\
  & \sum_{i=1}^{n/2} \lambda_{ji}r_i^2\omega_i^{k+j}\sin(\omega_i\alpha+
    \frac{\pi k}{2})\sin(\omega_i\alpha+\frac{\pi j}{2}) \nonumber \\
  &= c_{kj1}\cos\left(\frac{\pi}{2}(k-j)\right) \quad .
\end{align}
Observing that $\cos\left(\frac{\pi}{2}(k-j)\right)=0$ for $k-j$ odd and $\pm 1$
for $k-j$ even yields the result in (\ref{equ:inner_product_relation}).
\end{IEEEproof}

\begin{lemma}
For a set of arbitrary real diagonal matrices
 $\Lambda_k=\mbox{diag}(\lambda_{k,1},\lambda_{k,2},\cdots,\lambda_{k,n/2})$
 for $k=1,2,\cdots,n$, the following relationship holds between the curve in
 (\ref{equ:parametric_curve}) and the $k$-th Frenet frame:
\begin{equation}
 e_k(\alpha)=\Lambda_k x_{\theta}^{(k)}(\alpha).
\label{equ:frenet_frame_curve_relation}
\end{equation}
\label{lemma:frenet_frame_curve_relation}
\end{lemma}

\begin{IEEEproof}
We couple the recursive generation of the Frenet frame vectors in
(\ref{equ:frenet_frame}) with mathematical induction to prove
(\ref{equ:frenet_frame_curve_relation}). In the sequel, for clarity, we drop
the curve parameter $\alpha$ and write $\x(\alpha)$ as $\x$. We
initialize the induction by computing the first two Frenet frames:
$\e_1 = \frac{\x'}{\norm{\x'}}=c_{110}^{-1} \x'$,
where $\Lambda_1=c_{110}^{-1}\mathbf{I}$.
Then, for $k=2$, we have
\begin{equation}
\tilde{\e}_2 = \x^{(2)}- \langle \x^{(2)},\e_1 \rangle \e_1 
 = \x^{(2)}- \frac{\langle \x^{(2)},\x^{(1)} \rangle}{c_{110}} 
\x^{(1)}
 = \x^{(2)} ,
\end{equation}
where the last equality follows from Lemma \ref{lemma:inner_product_relation}.

Then, $\e_2=\Lambda_2\x^{(2)}$ where 
$\Lambda_2=c_{220}^{-1}\mathbf{I}$ since $c_{110}=1$. For $k\geq 3$, we get
\begin{align}
\tilde{\e}_k &= \x^{(k)}-\sum_{i=1}^{k-1}
 \langle \x^{(k)},\e_i \rangle \e_i 
 \stackrel{(i)}{=} \x^{(k)}-\sum_{j\in\mathcal{A}_k} 
 \langle \x^{(k)}, \Lambda_j \x^{(j)} \rangle 
  \Lambda_j \x^{(j)} \nonumber \\
& \stackrel{(ii)}{=} \x^{(k)}-\sum_{j\in\mathcal{A}_k}c_{kj1}\Lambda_j\x^{(j)} 
 \stackrel{(iii)}{=} \x^{(k)}-\Gamma_k\x^{(k)},
\end{align}
where $\mathcal{A}_k=\{l=1,2,\cdots,\lfloor (k-1)/2 \rfloor \}$ is the set
of positive even integers less than $k$ for $k$ even or set of
positive odd integers less than $k$ for $k$ odd; and $\Gamma_k$ is
another diagonal matrix such that $\mbox{diag}(\Gamma_k)$ is the
resultant vector in the direction of $\x^{(k)}$.  We then have that
(\emph{i}) follows from recursion on $\e_k$, (\emph{ii}) follows from Lemma
\ref{lemma:inner_product_relation}, and (\emph{iii}) follows from linear
superposition of vectors in a common basis (since the same basis results
after an even number of differentiations of sin/cos pairs).
\end{IEEEproof}

The local curvature (first generalized curvature) is obtained directly from
(\ref{equ:frenet_frame_curve_relation}) or as reciprocal of the limiting
case of the circumradius function as $\alpha_1 \rightarrow \alpha_2$.
The global circumradius, which defines our tube radius, is simply
\begin{equation}
\rho_G=\min_{\alpha_1,\alpha_2\in I}\rho(\alpha_1,\alpha_2)\quad . 
\label{equ:global_circumradius}
\end{equation}
Note that this global circumradius is equivalent to the maximal small ball radius defined in \cite[Definition 2]{VaishampayanCosta2003}. The minimum distance of the code (defined to be the smallest distance
between folds of the curve) is then simply twice the global circumradius. 

Squaring (\ref{equ:circumradius}) and eliminating the sine term yields
\begin{align}
& \rho^2(\alpha_1,\alpha_2) \nonumber\\
&= 
  \frac{1}{4} \frac{ \norm{x_{\theta}(\alpha_1)-x_{\theta}(\alpha_2)}^4 
  \norm{x_{\theta}^\prime(\alpha_2)}^2} { \left( \norm{x_{\theta}(\alpha_1)-x_{\tt}(\alpha_2)}^2 
      \norm{x_{\theta}^\prime(\alpha_2)}^2 %\right. \nonumber \\
%& \left. 
-\langle x_{\theta}(\alpha_1)-
      x_{\theta}(\alpha_2),x_{\theta}^\prime(\alpha_2)\rangle^2 \right)}, %^{-1},
\label{equ:circumradius_sq}
\end{align}
Define $\Delta=\alpha_1-\alpha_2$. Then, substituting (\ref{equ:parametric_curve}) into (\ref{equ:circumradius_sq}) produces
\begin{align}
\rho_{\theta}^2(\Delta)& =\frac{t_1^2t_2}{4t_1t_2-4t_3^2}
\end{align}
where
\begin{align}
t_1 &=  
    \begin{dcases}
        2\sum_{i=1}^{n/2}r_i^2(1-\cos(\omega_i\Delta)), &\text{n even,} \\
        2\sum_{i=1}^{(n\num{-1})/2}r_i^2(1-\cos(\omega_i\Delta)) + b^2\Delta^2, &\text{n odd,}
    \end{dcases}    \nonumber \\
t_2 &=  
    \begin{dcases}
        \sum_{i=1}^{n/2} \omega_i^2 r_i^2, &\text{n even,} \\
        \sum_{i=1}^{(n\num{-1})/2} \omega_i^2 r_i^2 + b^2, &\text{n odd,}
    \end{dcases}    \nonumber
\end{align}
and
\begin{align}    
t_3 &=  
    \begin{dcases}
        \sum_{i=1}^{n/2}r_i^2\omega_i\sin(\omega_i\Delta), & \text{n even,} \\
        \sum_{i=1}^{(n\num{-1})/2}r_i^2\omega_i\sin(\omega_i\Delta) + b^2\Delta, &\text{n odd.}
    \end{dcases}   
    \label{equ:rho_theta_sq}
\end{align}
The global circumradius function of (\ref{equ:global_circumradius}) then becomes
\begin{align}
\rho_{G}(\tt)=\min_{\Delta\in [0,2\pi]}\rho_{\tt}(\Delta) \quad .
\label{equ:rho_theta_global}
\end{align}
The aforementioned analysis shows that constant curvature curves have the nice property that the circumradius
is only a function of the difference between two points on the curve.  It may also be verified via L'H\^{o}pital's rule that $\rho(0)=\chi_1^{-1}$.

\subsection{Optimization of C3T encoder}
\label{sec:optimization}

While focusing on optimizing the C3T curves for the low \textrm{SNR} regime, we use the lowest unique harmonic frequencies for $\omega_j$ terms in
$\theta=\{r_1,r_2,\cdots,r_{n/2},\omega_1,\omega_2,\cdots,\omega_{n/2}\}$, thus letting $\omega_j=j$, for all $j=1,2,\cdots,n/2$. Note that this differs from the formulation in  \cite{VaishampayanCosta2003}, which uses equal radii and a sequence of frequency parameters $\omega_j = a^j$, $\forall j=0,1,\cdots,n/2-1$. This is called an exponential strut \cite{Sloan2010}, wherein the frequency parameters are optimized for a constant radius chosen to satisfy a power constraint. This is directly related to the \emph{fat strut} optimization \cite{Sloan2009} that involves an $n$-dimensional cylinder with a spherical cross-section. Here, the end-points are anchored to lattice points in $\mathbb{Z}^n$ that do not have any lattice points within the cylinder. Two fat struts of equal length may have different radii. Therefore, the optimization problem is to find end-points that maximize the radius. The resulting fat strut, thus, is a cylinder with maximized volume \cite{Sloan2011}. In \cite{Campello2013}, a strategy involving fat struts is used for designing a set of spherical curves on flat tori, each one of constant curvature that for a given radius provides a larger total length. A related problem is the cylinder optimization of \cite{Horvath1970}, which involves lattice sphere packing in three or higher dimensions with an infinite cylinder that does not touch any of the spheres.

With the decision to use the lowest harmonic frequencies, all that is left to optimize then in $\theta$ are the radii parameters $\mathbf{r}=\{r_1,\cdots,r_{n/2}\}$. Following \cite{Kravtsov2007}, we know that packing density plays a key role in the performance of analog codes. We select the parameters of our encoder functions to maximize the density of the (hyper)-tube contained within the $n$-sphere such that the packing density is 
\begin{align}
  \mbox{density}&=\frac{\mbox{path length}\times(\mbox{tube
    cross-sectional volume})}{\mbox{Volume of bounding n-sphere}} \nonumber \\
  &=\frac{L_{\mathbf{r}} \mathcal{C}_{n\num{-1}}\rho_{G,\mathbf{r}}^{n\num{-1}}}{\mathcal{C}_n(1+\rho_{G,\mathbf{r}})^n},
  \label{equ:density}
\end{align}
where we have used the result from Lemma~\ref{lemma:hypervol}.  Therefore, from (\ref{equ:density})
and utilizing (\ref{equ:L}), (\ref{equ:rho_theta_global}), and
(\ref{equ:V_n(R)}), we define our tube-packing objective function for the C3T encoder map as
\begin{align}
  J(\mathbf{r})&=\frac{L_{\mathbf{r}}\rho_{G,\mathbf{r}}^{n\num{-1}}}{(1+\rho_{G,\rr})^n}, \label{equ:J}
  \end{align}
where
\begin{align}
  \rho_{G,\mathbf{r}} &= \sqrt{\min_{\Delta\in [0,2\pi]}\rho_{\mathbf{r}}^2(\Delta)},\label{equ:tube_radius}\\
  \rho_{\mathbf{r}}^2(\Delta)& =\frac{[\sum_{i=1}^{n/2}r_i^2(1-\cos(i\Delta))]^2}
      {2\sum_{i=1}^{n/2}r_i^2(1-\cos(i\Delta))-\frac{[\sum_{i=1}^{n/2}ir_i^2\sin(i\Delta)]^2}{\sum_{i=1}^{n/2}i^2r_i^2}},
  \label{equ:sq_radius}
\end{align}
and
\begin{align}         
  L &=   2\pi \sqrt{\sum_{i=1}^{n/2} i^2 r_i^2}.
\end{align}

Since closed-form gradients do not exist for the objective function $J(\mathbf{r})$, we use the SPSA algorithm ~\cite{Spall1992} to optimize
\begin{align}
\hat{\rr} = \max_{\rr \in \mathcal{S}^{n/2}} J(\rr),
\label{equ:spsa_optimizer}
\end{align}
where $\mathcal{S}^{n/2}$ is the $n/2$-dimensional unit sphere. The SPSA algorithm is a stochastic gradient approximation method capable of finding global minima requiring only two measurements of the objective function per iteration regardless of the dimension of the optimization problem. We employ the iteration
\begin{align}
  \rr_{k+1}&\leftarrow\rr_k + a_k \hat{\gg}_k(\rr_k), \nonumber \\
  \rr_{k+1}&\leftarrow\frac{\mathbf{r_{k+1}}}{\norm{\mathbf{r_{k+1}}}} ,
  \label{equ:spsa_iterate}
\end{align}
where $\mathbf{r_k}$ is the $k$-th iterate, $\hat{\gamma}_k(\mathbf{r_k})$ is the
estimate of the gradient of the objective
$\gamma(\mathbf{r})=\partial J(\mathbf{r})/\partial \mathbf{r}$, and $\{a_k\}$ is a positive number sequence converging to 0.  The $j$-th component of
the stochastic perturbation gradient estimator is
\begin{align}
  (\hat{\gamma}_k(\mathbf{r_k}))_j = \frac{J(\mathbf{r_k}+c_k\DD_k)-J(\mathbf{r_k}-c_k\DD_k)}{2c_k(\DD_k)_j}.
  \label{equ:spsa_gradient_estimate}
\end{align}

We set the random perturbation vector $\DD_k$ according to the Rademacher distribution, that is, $\DD_k$ is drawn from a Bernoulli distribution $\pm 1$ with equal probability. In our experiments, we set $a_k=0.01/(k+11)^{0.602}$ and $c_k=0.01/(k+1)^{0.101}$, which are adjustable and key parameters for tuning the SPSA performance. We iterate until the deviation between consecutive cost function evaluations is less than a tolerance level of $10^{-3}$.  In order to achieve a more optimal choice of radii, we improved the SPSA algorithm by restarting its operation after a certain period of time using the radii obtained up until that point. Algorithm \ref{alg:spsa} summarizes this procedure.

\begin{comment}
%-----------------------------------------------------------------------------------
\begin{algorithm}[htbp!]
 1. Set max density to $0$\;
 2. Initialize radii to $r=\{1,1,\dots,1\}_{n/2}$ and normalize\;
 3. Pre-compute $\{\textrm{cos},\textrm{sin}\}$-vectors for $\Delta = [0,\pi]$\;
 \For{count = 1 to 600}
 {
  \While{consecutive cost deviation $>$ tolerance, $10^{-3}$}
  {
   4. Draw values from the Rademacher distribution\;
   5. Compute objective function\;
   6. Stochastic perturbation gradient estimator\;
   7. Update the radii vector\;
  }
  8. Compute density using current radii vector\;
  \If{current density $>$ max density}
  {
     9. max density := current density
  } 
 } % for
 \KwResult{Print global circumradius, tub density and the optimized radii}
 \caption{\textcolor{red}{Improved SPSA iterations.}}
 \label{alg:spsa}
\end{algorithm}
%-----------------------------------------------------------------------------------
\end{comment}

\begin{figure}[t]
\centering
    \includegraphics[width=1.0\columnwidth]{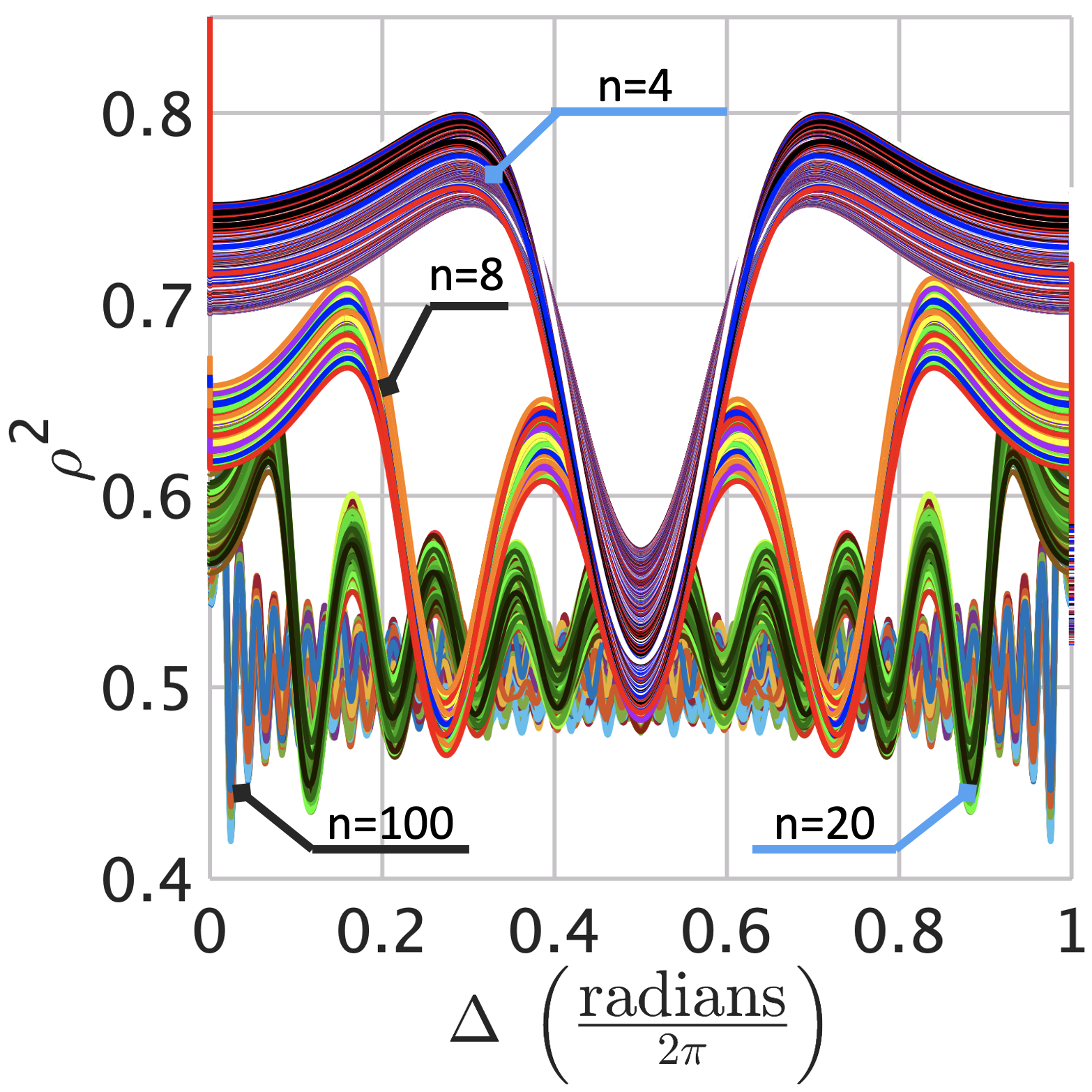}
    \caption{Tube radius as a function of the distance between two points on the curve. A group of curves indicates 600 iterations each for even dimensions: $n=4$ (top), $8$ (second group from top), $20$ (third group from top), and $100$ (bottom). The smallest circumradius for $n=4$ is found at $\Delta = \pi$. As the dimension grows, the smallest circumradius is found at successively smaller $\Delta$.
    }
    \label{fig:circum_even}
\end{figure}

\begin{figure}[t]
\centering
    \includegraphics[width=1.0\columnwidth]{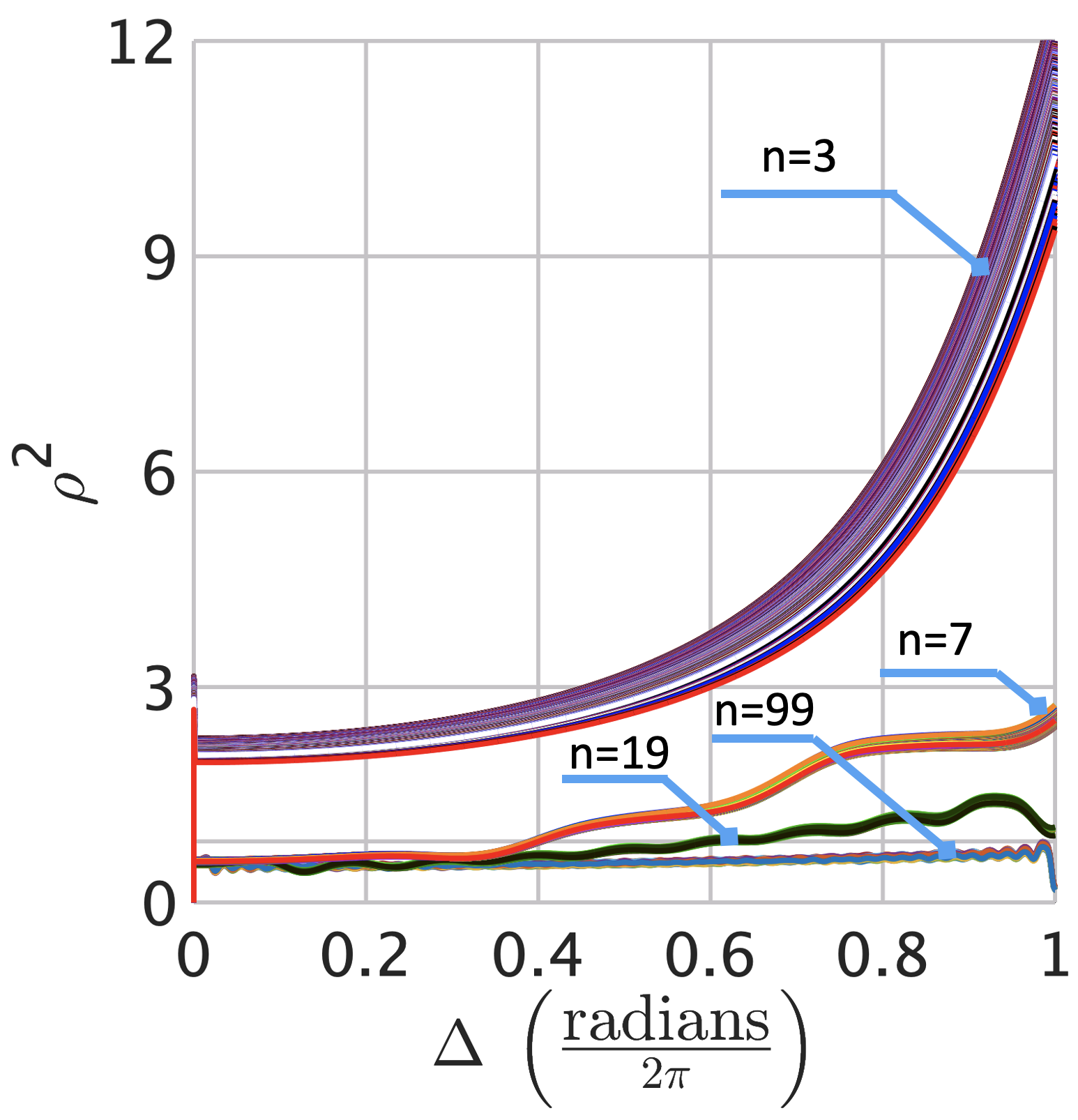}
    \caption{As in Fig.~\ref{fig:circum_even} but for odd dimensions: $n=3$ (top), $7$ (second group from top), $19$ (third group from top) and $99$ (bottom).
    }
    \label{fig:circum_odd}
\end{figure}

%-----------------------------------------------------------------------------------
    \begin{algorithm}[H]
    \caption{Improved SPSA.}\label{alg:spsa}
    \textbf{Input:} Decoder dimensionality $n$\\
    \textbf{Output:} Optimized radii $r_{opt}$ 
    \begin{algorithmic}[1]
    %\Require $n \geq 0$
    %\Ensure $y = x^n$
    \State $\mathbf{r} \gets \{1,1,\dots,1\}_{n/2}$ 
    \State $\mathbf{r_1} \gets \mathbf{r} / \lVert \mathbf{r} \rVert$
    \State $Max \gets 0$            \Comment{Max density}
    \State $Pre \gets \infty$       \Comment{Previous objective}
    \State $Dev \gets \infty$       \Comment{Objective deviation}
    \State $\{\textrm{cos},\textrm{sin}\}$ for $\Delta = [0,\pi]$ \Comment{Init $\{\textrm{cos},\textrm{sin}\}$-vectors}
    \For{it = 1:600}
        \State $k \gets 0$
        
        \While{$Dev > 10^{-3}$}
            \State $k \gets k+1$
            \State $\DD_k \gets \mathcal{B}(p=0.5)$     \Comment{Bernoulli distribution}
            \State $\hat{\gamma}_k(\mathbf{r_k}) \gets \frac{J(\mathbf{r_k}+c_k\DD_k)-J(\mathbf{r_k}-c_k\DD_k)}{2c_k(\DD_k)_j}$ \Comment{Estimate gradient}
            \State $\rr_{k+1} \gets \rr_k + a_k \hat{\gg}_k(\rr_k)$
            \State $\rr_{k+1} \gets \frac{\mathbf{r_{k+1}}}{\norm{\mathbf{r_{k+1}}}}$
            \State $J(\mathbf{r}) \gets \frac{L_{\mathbf{r}}\rho_{G,\mathbf{r}}^{n-1}}{(1+\rho_{G,\rr})^n}$     \Comment{Objective function}
            \State $Dev \gets \lVert Pre - J(\mathbf{r}) \lVert_1 $
            \State $Pre \gets J(\mathbf{r})$
        \EndWhile
        \State $\rho_{\mathbf{r}}^2(\Delta) \gets \frac{[\sum_{i=1}^{n/2}r_i^2(1-\cos(i\Delta))]^2}{2\sum_{i=1}^{n/2}r_i^2(1-\cos(i\Delta))-\frac{[\sum_{i=1}^{n/2}ir_i^2\sin(i\Delta)]^2}{\sum_{i=1}^{n/2}i^2r_i^2}}$
        \State $\rho_{G,\mathbf{r}} \gets \sqrt{\min_{\Delta\in [0,\pi]}\rho_{\mathbf{r}}^2(\Delta)}$           \Comment{Global circumradius}
        \State $L \gets 2\pi \sqrt{\sum_{i=1}^{n/2} i^2 r_i^2}$                                                           \Comment{Curve length}
        \State $Den \gets \frac{L \; \mathcal{C}_{n-1}\rho_{G,\mathbf{r}}^{n-1}}{\mathcal{C}_n(1+\rho_{G,\mathbf{r}})^n}$ \Comment{Tube density}
        \If{$Den > Max$}
            \State $Max \gets Den$
            \State $r_{opt} \gets \rr_{k+1}$
        %\ElsIf{$N$ is odd}
        %    \State $y \gets y \times X$
        %    \State $N \gets N - 1$
        \EndIf
        \State $\mathbf{r_1} \gets r_{opt}$ \Comment{Start with the best radii}
    \EndFor
    \State \Return $r_{opt}$
    \end{algorithmic}
    \end{algorithm}
    %-----------------------------------------------------------------------------------
For the range of $\Delta$-values in \eqref{equ:tube_radius}, we picked a sufficiently small resolution of $10^{-4}$ to reliably find the minima shown in Figs. \ref{fig:circum_even} and \ref{fig:circum_odd}. Here, the tube radius is plotted as a function of $\Delta$ as it evolves over $600$ SPSA iterations. Note that \eqref{equ:sq_radius} is symmetrical about $\pi$ and, therefore, the search space for the minimum is reduced to $[0,\pi]$.  For $n=4$, the smallest tube radius is found for $\Delta = \pi$. We observe that, for even $n$, the smallest tube radius is found at a smaller $\Delta$ as the dimensionality increases. %Therefore, the SPSA algorithm does not work for $n=2$. 
This is also true for odd $n$.
%-----------------------------------------------------------------------------------
%-----------------------------------------------------------------------------------

%-----------------------------------------------------------------------------------
%-----------------------------------------------------------------------------------

Table~\ref{tab:radii_params} shows the optimizing radii vector $\hat{\mathbf{r}}$, the global circumradius (tube radius) $\rho_{G,\mathbf{r}}$, and the tube density from (\ref{equ:density}) for a small sample of even- and odd-dimensional C3T codes computed from SPSA optimization of (\ref{equ:J}).  Similar to previous results on sphere packing ~\cite{ConwaySloane1998}, the tube packing density decreases rapidly with an increase in dimensions. Table~\ref{tab:nonharmonic} lists $\rho$ for various choices of frequency parameters for $n=6$.  In particular, the frequency parameter combination of $\{1,3,5\}$ results in a higher tube density than the $\{1,2,3\}$.
%-----------------------------------------------------------------------------------
\begin{table}[t]
    \centering
    \caption{Optimized encoder parameters for C3T codes}
    \begin{threeparttable}
    \begin{tabular}{cccc}
      \toprule
      $n$ & $\theta =\{[\mathbf{r}], n\mbox{ even or } [\mathbf{r},b], \; n\mbox{ odd}\}$\tnote{a} & $\rho_{G,\mathbf{r}}$\tnote{b} & \mbox{Tube Density}\tnote{c} \\
      \midrule
        3 &  [0.5000, 0.8660] & 2.0000 &  0.6981 \\
        4 &  [0.8165, 0.5774] & 0.8339 & 0.3783 \\
        5 &  [0.01490, 0.7810, 0.6244] & 0.9058 & 0.2653 \\
        6 &  [0.5835, 0.6463, 0.4918] & 0.7614 & 0.1122\\
        7 &  [0.2185, 0.6976, 0.4431, 0.5188] & 0.8273 &  0.0650 \\
        8 &  [0.5125, 0.5162, 0.5551, 0.4035] & 0.7541 & 0.0293\\
        9 &  [0.3369, 0.6007, 0.4652, 0.4053, 0.3807] & {red}{0.7748} &  {red}{0.0144} \\
        10 &  [0.4505, 0.4402, 0.4708, 0.500, 0.3628] & 0.7442 & 0.0070\\
        11 &  [0.3884, 0.4565, 0.4431, 0.3809, 0.3785, 0.3951] & {red}{0.7217} & 0.0025 \\
        12 & [0.4141, 0.3953, 0.4102, 0.4346, 0.4566, 0.3265] & 0.7387 & 0.0016 \\
      \bottomrule
    \end{tabular}
    \begin{tablenotes}[para]
    \item[a]The optimized value of radii vector $\hat{\mathbf{r}}$ were computed using SPSA for a small sample of even-dimensional C3T codes.  
    \item[b]The global circumradius (tube radius) $\rho_{G,\mathbf{r}}$ as in \eqref{equ:tube_radius}
    \item[c]The tube density as in (\ref{equ:density}) computed from the optimization of (\ref{equ:J}) according to (\ref{equ:spsa_iterate}) and (\ref{equ:spsa_gradient_estimate}).
    \end{tablenotes}
    \end{threeparttable}    
    \label{tab:radii_params}
\end{table}
%-----------------------------------------------------------------------------------
%-----------------------------------------------------------------------------------
\begin{table}[t]
\caption{\label{tab:nonharmonic} {red}{Tube densities for $n=6$.}}
\centering
\begin{threeparttable}
\begin{tabular}{c|c}
$\{\omega_1,\omega_2,\omega_3\}$ & $\rho$\\
\hline\hline
$\{1,2,3\}$\tnote{a} & $0.1122$\\
$\{1,2,5\}$ & $0.1127$\\
$\{1,3,5\}$ & $0.1219$\\
$\{1,1,1\}$ & $0.09921$\\
$\{1,1,3\}$ & $0.08071$\\
$\{1,3,3\}$ & $0.08069$\\
$\{1,4,5\}$ & $0.06285$\\
\hline
\end{tabular}
\begin{tablenotes}
    \item[a] These are also the parameters of the harmonic BW expansion used in this paper.
\end{tablenotes}
\end{threeparttable}
\end{table}
%-----------------------------------------------------------------------------------

\subsection{Analysis}
\label{subsec:perf}
%\subsection{C3T Encoder}
%\label{sec:c3t_encoder}
The functional form of our C3T encoder maps $\mathbf{x}$ can be solved through
the Frenet-Serret system of differential equations in
(\ref{equ:frenet-serret}).  This is true whether $n$ is even or odd.
Thus far we have focused only on the even $n$ case.  When $n$ is odd
the C3T code becomes a generalized helix and no longer has an encoder
locus on the $n/2$ flat torus.  However, the same general technical approach
to optimizing the encoder map applies to odd-dimensional spaces. The cost function of (\ref{equ:J}) is applicable for odd-dimensional codes except that the expression for the circumradius function is now obtained by plugging
(\ref{equ:parametric_curve2}) into (\ref{equ:circumradius_sq}).
For example, for $n=3$, we get
\begin{equation}
  \rho^2(\Delta) = \frac{\frac{1}{4}t_1^2(b^2+r^2)}{t_1(b^2+r^2)-t_2^2},
  \label{equ:rho_squared_3Dhelix}
\end{equation}
where
\begin{align}
t_1 = b^2\Delta^2+2r^2(1-\cos\Delta),\;t_2 = b^2\Delta+r^2\sin\Delta.
\label{equ:t_definitions}
\end{align}

\begin{figure}[t]
\begin{minipage}[b]{.24\linewidth}
  \centering
  \centerline{\includegraphics[width=\columnwidth]{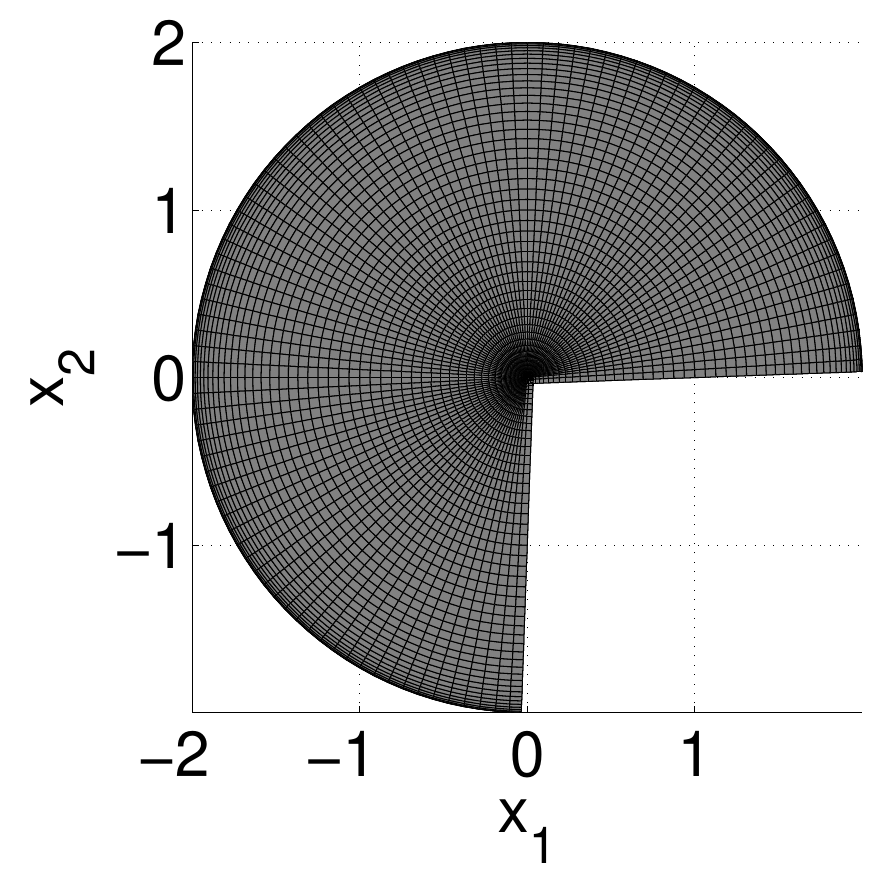}}
  \centerline{(a) 2-D ribbon}\medskip
\end{minipage}
\hfill
\begin{minipage}[b]{0.32\linewidth}
  \centering
  \centerline{\includegraphics[width=\columnwidth]{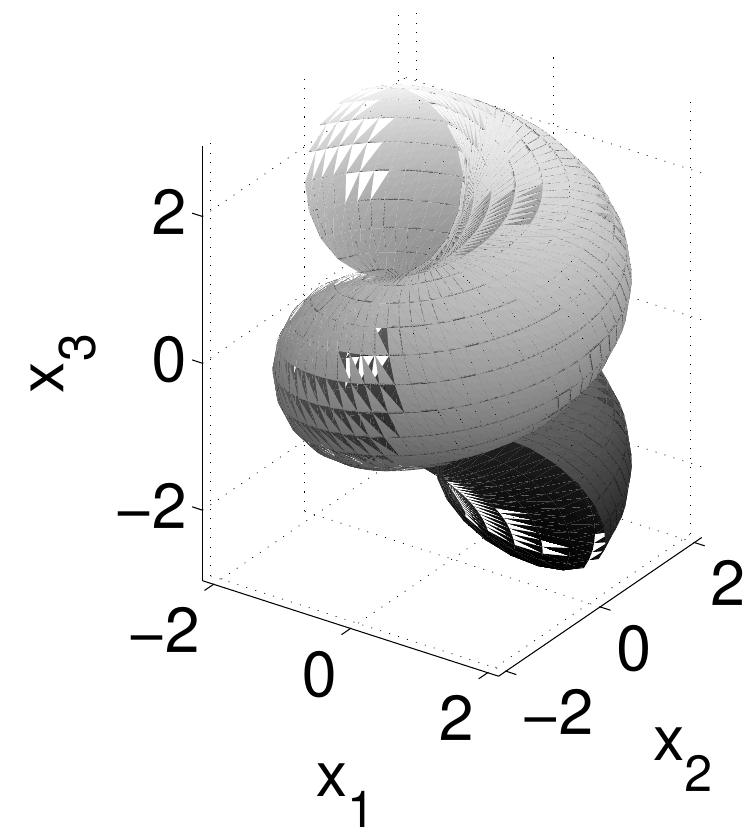}}
  \centerline{(b) 3-D tube}\medskip
\end{minipage}
\hfill
\begin{minipage}[b]{0.36\linewidth}
  \centering
  \centerline{\includegraphics[width=\columnwidth]{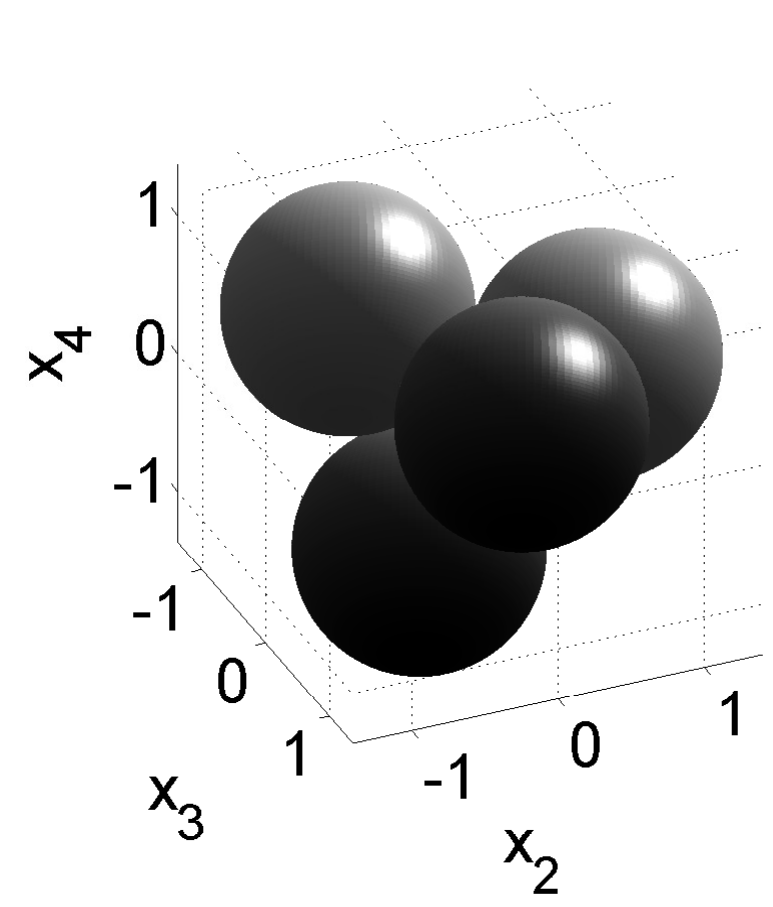}}
  \centerline{(c) 4-D tube cross-section}\medskip
\end{minipage}
\caption{Dense tube packings for $n=2$, $3$, and $4$. For $n=4$, the plot is the intersection of the tube in 4-D with 3-D hyperplane through the origin.}
\label{fig:ribbon_tube_balls}
\end{figure}

The local curvature (first generalized curvature) is obtained
directly from (\ref{equ:frenet_frame_curve_relation}) or as the reciprocal of
the limiting case of the circumradius function when $\alpha_1\rightarrow\alpha_2$.  From L'H\^{o}pital's rule, the
circumradius function collapses back to the inverse local radius of
curvature function $\chi_1^{-1}$ from
(\ref{equ:frenet_frame_curve_relation}) such that $\lim_{\Delta\rightarrow
  0} \rho(\Delta)=\chi_1^{-1}$.  For example, $n=3$ yields the first
generalized curvature as $\chi_1=r/(r^2+b^2)$. On the other hand, for $n=4$, the
first generalized curvature is
$\chi_1=\sqrt{r_1^2\omega_1^4+r_2^2\omega_2^4}$. For constant
curvature curves with global circumradius $\rho_G$ (tube radius from
(\ref{equ:global_circumradius})), the minimal Euclidean distance
between any two points on the curve that involves jumping to another
section of the curve is $2\rho_G$. The constant curvature property
guarantees that any subset of points that we choose maintain
their same relative location to each other as we move each point along
the curve by the same path length. This is similar to the
geometrically uniform property of Forney \cite{Forney1991} but for
continuous-alphabet codes. The squared circumradius function in
(\ref{equ:rho_theta_sq}) is even symmetric. In Fig. \ref{fig:ribbon_tube_balls}, we show surface plots of the computed tube packings. For $n=4$, we plot the 3-D cross-section of the 4-D tube packing. Interestingly, the
cross-sections of each of $n$ dimensional tubes tend to form
sphere packings in $n\num{-1}$ dimensional cross-sectional Euclidean
space.

%-----------------------------------------------------------------------------------
%-----------------------------------------------------------------------------------

There are many interesting special cases that emerge when the
constant curvature curves are restricted to the radii vector
$\rr=[\sqrt{2/(n\num{-1})},\cdots,\sqrt{2/(n\num{-1})},\sqrt{1/(n\num{-1})}]$.
When the frequency vector is set to $\omega=[1,2,\cdots,n/2]$, the
resulting C3T curve intersected with a $n\num{-1}$-dimensional hyperplane
through the origin corresponds to the vertices of $(n\num{-1})$-simplices.
When the frequency vector is set to $\omega=[1,3,5,\cdots,n\num{-1}]$ the
resulting C3T-hyperplane slices are vertices of $(n\num{-1})$-orthoplexes
(cross-polytopes). For the $n=4$ case and letting $\omega=[1,4]$, we
get the vertices of the twisted cube in three dimensions.

\section{C3T Decoder}
\label{sec:decoder}
The maximum likelihood (ML) decoder for the encoder maps is
$\hat{\alpha}=\argmin_{\alpha\in I} \norm{y-x_{\hat{\theta}}(\alpha)}$,
which may have many local minima and requires a good initial guess. This decoder necessitates an exhaustive search over the 1-dimensional space $\alpha$ contained in interval $I$. Computing the natural logarithm of the posterior probability density function (pdf) as,
\begin{flalign}
  &\log p(\alpha|y)\nonumber\\
  &=
  \log[p(\alpha) \prod_{i=1}^{n/2}\mathcal{N}(y_{2i-1}; r_i\cos(\omega_i\alpha),\sigma^2)  \mathcal{N}(y_{2i}; r_i\sin(\omega_i\alpha),\sigma^2)] \nonumber\\
   &= -\frac{1}{2\sigma^2}\sum_{i=1}^{n/2}[(y_{2i-1}-r_i\cos(\omega_i\alpha))^2  +(y_{2i}-r_i\sin(\omega_i\alpha))^2 ] + c_1 \nonumber\\
  & = \frac{1}{\sigma^2}\sum_{i=1}^{n/2}r_i(y_{2i-1}\cos(\omega_i\alpha)+y_{2i}\sin(\omega_i\alpha))+c_2, 
\end{flalign}
it is evident that the ML decoder (also the MAP decoder for our uniform source case) reduces to,
\begin{align}
  \hat{\alpha}_{\textrm{MAP}} = \argmax_{\alpha\in [-\pi,\pi]} \sum_{i=1}^{n/2} r_i\left[ y_{2i-1}\cos(\omega_i\alpha)+y_{2i}\sin(\omega_i\alpha)\right] .
  \label{equ:map}
\end{align}
The ML decoder is generally intractable leading to a host of approximate decoding methodologies. While most even-$n$ constant curvature codes exploit the local isometry of the flat torus to a hyperrectangle as in \cite{VaishampayanCosta2003} to decode, here we go for a %closed-form 
direct function approximation using the neural network approach. 

The optimal decoder \cite{Hu2011} is the MMSE expressed as
\begin{align}
  \hat{\alpha}_{\textrm{MMSE}} 
  &=\argmin_{\alpha\in [-\pi,\pi]} \mathrm{E}\left\{\norm{y-x_{\hat{\theta}}(\alpha)}^2\right\} \nonumber\\
  &=\argmin_{\alpha\in [-\pi,\pi]} \mathrm{E}\left\{x_{\hat{\theta}}|y\right\} \nonumber\\
  &=\argmin_{\alpha\in [-\pi,\pi]} \int_{-\infty}^{\infty}
  x_{\hat{\theta}}(\alpha)p(x_{\hat{\theta}}(\alpha)|y) dx_{\hat{\theta}} \nonumber\\
  &=\frac{1}{p(y)} \argmin_{\alpha\in [-\pi,\pi]} \int_{-\infty}^{\infty} x_{\hat{\theta}}(\alpha)p(y|x_{\hat{\theta}}(\alpha))p(x_{\hat{\theta}}(\alpha))dx_{\hat{\theta}},
  \label{equ:mmse}
\end{align}
where 
\begin{align*}
    x_{\hat{\theta}}(\alpha) = 
    \begin{dcases}
        r_i \cos(\omega_i\alpha),\;i=1,\cdots, n/2,\\
         r_i \sin(\omega_i\alpha),\;i=1,\cdots, n/2,
    \end{dcases}
\end{align*}
The prior probability density function $p(x_{\hat{\theta}}(\alpha))$ follows from $\cos(u)$ or $\sin(u)$ of a uniformly distributed random variable $u \in \mathcal{U}[-1,+1]$ as %is
\begin{align*}
    p(x) = \frac{1}{\pi\sqrt{1+(x/r_i)^2}},\;x\in[-r_i,+r_i].
\end{align*}
The conditional probability density functions for the elements of the received vector $y$ given the elements of the transmitted vector $x_{\hat{\theta}}(\alpha)$ are
\begin{align*}
    p(y_{2i-1}|x_{2i-1}(\alpha))
    &= \frac{1}{\sigma\sqrt{2\pi}}\exp\left\{ -\frac{1}{2\sigma_w^2} (y_{2i-1}-r_i\cos(\omega_i\alpha))^2\right\},\\
    p(y_{2i}|x_{2i}(\alpha))
    &= \frac{1}{\sigma\sqrt{2\pi}}\exp\left\{ -\frac{1}{2\sigma_w^2} (y_{2i}-r_i\sin(\omega_i\alpha))^2\right\}.
\end{align*}
These probability density functions may also be found using numerical Monte Carlo methods \cite{Doucet2005}.

In particular, we train a five-layer MLP network using stochastic gradient descent with the Adam optimizer. The network uses a hyperbolic tangent activation function on all neurons.  For each hidden layer, the number of neurons gradually increases until the middle layer, from where the number of neurons gradually decreases for each additional layer (Table \ref{tab:mlp}). In this work, we explore gradually increasing dimensions together with the tangent hyperbolic activation function, \emph{tanh}, which has a superior performance over the standard sigmoid function. The output range of the \emph{tanh} function is $[-1, 1]$ and, unlike another popular activation function of rectified linear unit (ReLU) \cite{Farrell2021}, has zero mean.

%-----------------------------------------------------------------------------------
\begin{table}[t]
\caption{\label{tab:mlp} Size of the MLP network.}
\centering

\begin{tabular}{c|c|c|c|c|c|c}
Input & Layer 1 & Layer 2 & Layer 3 & Layer 4 & Layer 5 & Output \\ \hline
$n$ & $256$ & $512$  & $1024$ & $512$ & $256$ & $1$  
\end{tabular}
\end{table}
%-----------------------------------------------------------------------------------

The time complexity for training the MLP with a BW expansion factor of $n$ for $t$ epochs using $p$ examples is $\mathcal{O}(pt[nk+2k^2+8k^2+8k^2+2k^2+k]) = \mathcal{O}(pt[nk+20k^2]) = \mathcal{O}(ptk[n+20k])$, where $k$ is the dimension of the first and last hidden layers. The MLP inference time is $\mathcal{O}(k[n+20k])$. 
We compare this inference time with that of the non-MLP sub-optimal torus projection decoder in \cite{Campello2013}, wherein it is stated to be of the same complexity $\mathcal{O}(N\lVert u_i\rVert_1)$ as the torus decoder of \cite{VaishampayanCosta2003}; here, $N = n/2$ and $u_i = r_i\omega_i$. % in our notation. 
    That is, the number of $\{\textrm{cos},\textrm{sin}\}$-pairs in the encoder is $N$. For $n=8$, we have $r_i \approx 0.5$. Hence,

\begin{align*}
%\label{equ:vnc_complexity}
    N\lVert u_i \rVert_1
    &= n/2 \; r_i\lVert \omega_i\rVert_1 \\
    &= n/2 \; r_i\sum_{i=1}^{n/2} \left | \omega_i \right | \\
    &= n/2 \; r_i \frac{n/2(n/2+1)}{2} \\
    & \approx \frac{n^2(n+2)}{32}. 
\end{align*}
It follows that the complexity of the non-MLP decoders is $\mathcal{O}(n^3)$ or 2,600 times less than the C3T decoder. There exist off-the-shelf processors with dedicated hardware support for MLP processing. As a result, the complexity difference might not be as drastic in practical implementations of our MLP decoder. For a comparison between the accuracy of our torus projection decoder and the torus decoder in \cite{VaishampayanCosta2003}, we refer the reader to  Section \ref{sec:numexp}.

We train and test the neural network decoder with the following three decoder inputs.
\begin{itemize}
\item \textbf{Raw observations}
  Here, we just use the $n$ dimensional channel output as input to the decoder network.
\item \textbf{Torus projection (TP)}
As in ~\cite{Campello2013,VaishampayanCosta2003}, we exploit the
fact that the encoder locus lies on the $n/2$ flat torus and perform a
``torus projection'' as a feature extraction before decoding.  First, the $\ell_2$-norm is computed on the individual $\textrm{cos}/\textrm{sin}$ pairs in raw observations. Next, $\textrm{arccos}$ and $\textrm{arcsin}$ are applied to the individual $\textrm{cos}/\textrm{sin}$ terms, and the average of the angles derived; adding the angles reduces the uncertainty. The torus projection is obtained by reconstructing the $n$-dimensional vector using the known transmit radii $r_i$, known transmit frequencies $\omega_i$, and the above-mentioned derived angles.
\item \textbf{Angles only (AO)}
  Here, only the pairwise four-quadrant angles of the channel output are used as decoder inputs. This reduces the required number of neurons needed in the decoder network by a factor of two for a relatively small performance penalty in the low dimensional codes.
\end{itemize}

In low dimensions, we achieve accurate approximate decoders using only the estimated angles of the C3T-coded channel output. By marginalizing the radii parameters, we reduce the size of the decoder input by half. This greatly reduces the MLP size needed for decoding and may be useful for applications seeking very low decoder complexity.  The time complexity for training the Angle-Only MLP for $t$ epochs using $e$ examples is $\mathcal{O}(pt[nk/4+k^2/2+2k^2+2k^2+k^2/2+k/2]) = \mathcal{O}(pt[nk/4+5k^2]) = \mathcal{O}(ptk[n/4+5k])$, where $k/2$ is the dimension of the first and last hidden layers, that is, a complexity-reduction of at least four times over the methods using the full $n$-dimensional input.  The Angle-Only MLP inference time is $\mathcal{O}(k[n/4+5k])$. The following Theorem~\ref{theorem:alpha_given_y} derives the likelihood function for the angles-only decoder.
%In the next Section \ref{sec:numexp}, we plot the results of the C3T decoder with angle only inputs. 

\begin{theorem}
  Denote the polar coordinate transformation on our channel output $\mathbf{y}$ by $\eta_i=\mbox{atan2}(y_{2i},y_{2i-1})$ and
  $\beta_i = \sqrt{y_{2i-1}^2+y_{2i}^2}$, $i=1,2,\cdots,\frac{n}{2}$. Then, given $\alpha$, the (angles-only) likelihood function for $\ee$ is
\begin{align}
  p(\ee | \alpha) \propto \prod_{i=1}^{n/2}\left(1 - \sqrt{\pi}q_i(\alpha)\exp{q_i^2(\alpha)} \mbox{erfc}(q_i(\alpha))\right),
  \label{equ:alpha_posterior}
\end{align}
where $\textrm{erfc}(\cdot)$ is the complementary error function and
\begin{align}
  q_i(\alpha)=\frac{r_i}{\sqrt{2\sigma_w^2}}\cos(\eta_i-\omega_i\alpha).
\end{align}
\label{theorem:alpha_given_y}
\end{theorem}
\begin{IEEEproof}
See Appendix~\ref{sec:theorem_alpha_given_y_proof}.
\end{IEEEproof}

\vspace{0.5cm}

The following approximation \cite{Oldham2009}
\begin{align}
  \sqrt{\pi}xe^{x^2}\mbox{erfc}(x) &\approx \frac{2}{1+\sqrt{1+2x^{-2}\Phi(x)}},
\end{align}
  where $\Phi(x) = 1 - c \exp(-xP(x))$, $P(x) = a\left[1-bx^2(1-\frac{a}{\pi^2}x)\right]$, $c = 1-\frac{2}{\pi}$, $a=0.8577$, and $b=0.024$, leads to the posterior in (\ref{equ:alpha_posterior}) become
\begin{align}
  p(\ee | \alpha) & \approx \kappa \prod_{i=1}^{n/2} \frac{g(q_i)-1}{g(q_i)+1},
\end{align}
where
\begin{align}
  g(q_i) & = \sqrt{1+2q_i^{-2}-2cq_i^{-2}\exp(-aq_i+abq_i^3-\frac{a^2b}{\pi^2}q_i^4)},
\end{align}
and
\begin{align}
  q_i & = \frac{r_i}{\sqrt{2\sigma_w^2}}\cos(q_i-\omega_i\alpha). 
\end{align}

\section{Numerical Experiments}
\label{sec:numexp}
We optimized (\ref{equ:J}) for $n = 2, 3, 4, 5, 6, 7, 8, 9, 10, 20, 40,$ and $100$ based on the SPSA encoder maps and then trained the MLP decoder networks for each type of feature set and for dimensions $n = 2, 4, 6, 8, 20, 40$, and $100$. The training was carried out over 50-100 epochs using 15,000 examples each at \textrm{SNR} levels of $\minus5$, $0$, $+5$, and $+10$ dB. MLP decoders were also trained for $n = 2$ and $n = 3$ at \textrm{SNR} levels $0$, $+7$, $+15$, and $+20$ dB in order to facilitate comparisons with other $1$:$2$ and $1$:$3$ BW expansion coders. To enable comparisons with spherical codes, an MLP decoder was trained for $n = 8$ at \textrm{SNR} levels $+10$, $+15$, $+20$, and $+25$ dB.  Some networks needed less than 50 epochs to converge while others required more than 50 epochs. The source symbols $s$ were drawn from $s \sim \mathcal U[-1,+1]$ and then stretched using $\alpha(s) = \beta \pi s$, where $\beta=0.9$ was used to introduce some guard against the end points of the curve.  The learning rates used were in the range from $1e\num{-6}$ to $1e\num{-4}$ and optimally selected to minimize the mean-squared error for each combination of dimension and feature type. In order to optimize the training, we developed an adaptive learning rate summarized in Algorithm \ref{alg:adaptive}. The Adam optimizer had a decay setting of $1e\num{-6}$ for all training instances. Once trained, the networks were evaluated at \textrm{SNR} levels from $\minus10$ dB to $+10$ dB using examples drawn from the same distribution as above.

%-----------------------------------------------------------------------------------
\begin{algorithm}[H]    
\caption{Optimized training procedure with adaptive learning rate.}\label{alg:adaptive}
    \textbf{Input:} Number of epochs $N$, initial learning rate $lr$, weight decay $dcy$\\
    \textbf{Output:} Trained MLP weights $wts$
    \begin{algorithmic}[1]
    \State $opt \gets \Call{Adam}{lr,dcy} $           \Comment{Adam optimizer}
    \State $cnt \gets 0$                                \Comment{Loss counter}
    \State $wts \gets \mathcal{U}(-1,+1)$               \Comment{Initialization of weights}
    \For{epoch = 1:N}
        \State $wts,loss \gets \Call{Train}{wts,opt}$    \Comment{Train network}
        \If{epoch == 1}
            \State $ pre \gets loss $
        \EndIf
        \State $cnt \gets cnt + 1$
        \State $ ave \gets (pre-loss) / cnt $         \Comment{Average change in loss}
        \State $ rel \gets \lVert ave / loss \lVert_1 $ \Comment{Relative change in loss}
        \If{$(cnt \geq 8)  \And  (rel < 0.1) $}
            \State $pre \gets loss$
            \State $lr \gets lr/2 $                     \Comment{Set new learning rate}
            \State $opt \gets \Call{Adam}{lr,dcy} $
            \State $cnt \gets 0$ 
        \EndIf    
    \EndFor
    \State \Return $wts$ 
    \end{algorithmic}
    \end{algorithm}
    %-----------------------------------------------------------------------------------

\begin{figure*}
\centering
    \includegraphics[width=\linewidth]{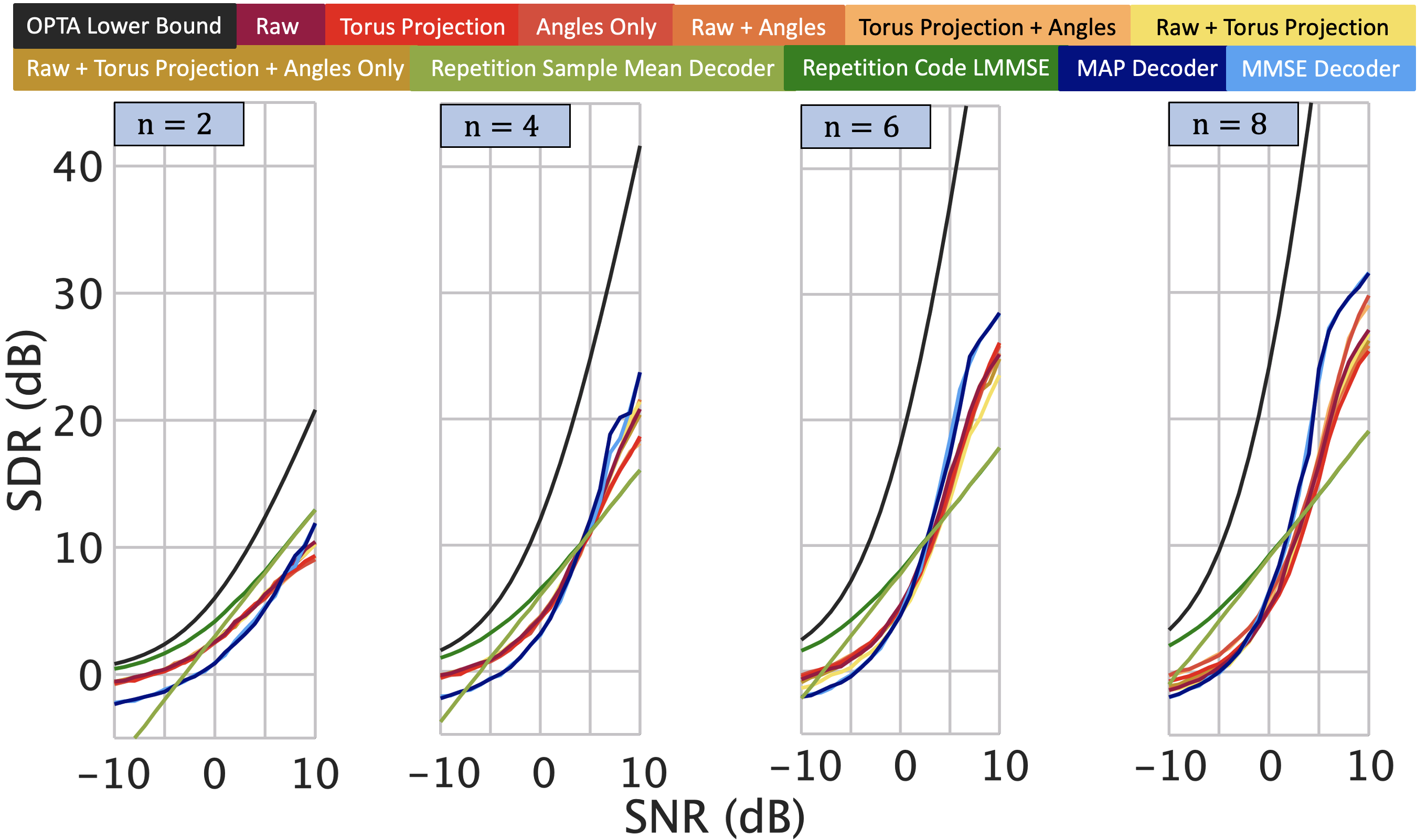}
    \caption{Performance of the proposed C3T codes ($n = 2, 4, 6$, and $8$) compared with the repetition code sample mean decoder, the repetition code optimal LMMSE decoder, the} MAP decoder, and the OPTA lower bound.
    \vspace{-0.5cm}
    \label{fig:perf_n2n4n6n8}
\end{figure*}

\begin{figure*}
\centering
    \includegraphics[width=\linewidth]{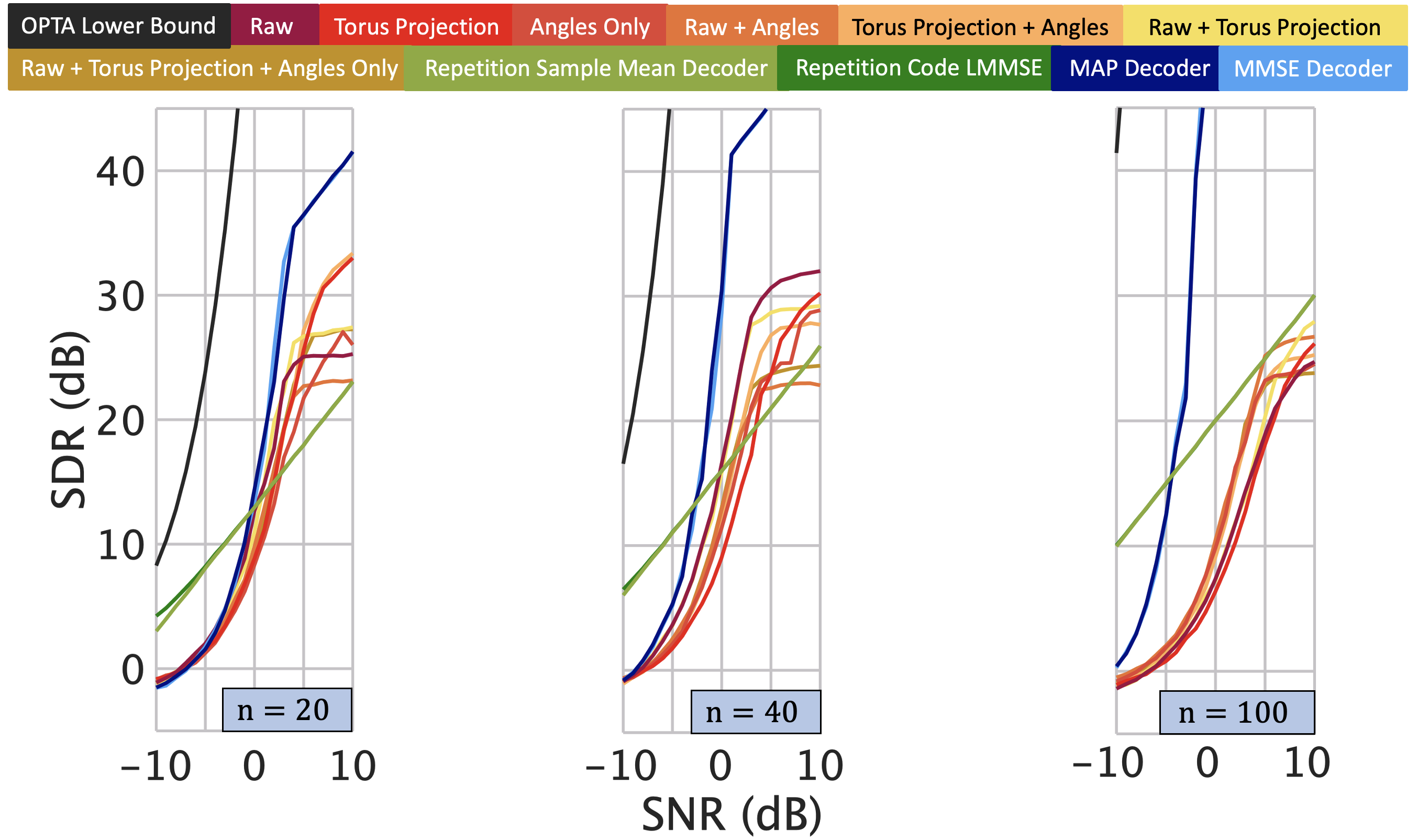}
    \caption{As in Fig.~\ref{fig:perf_n2n4n6n8}, but for C3T codes with $n = 20, 40$, and $100$.}
    \vspace{-0.5cm}
    \label{fig:perf_n20n40n100}
\end{figure*}

\begin{figure}[t]
\centering
    \includegraphics[width=1.0\columnwidth]{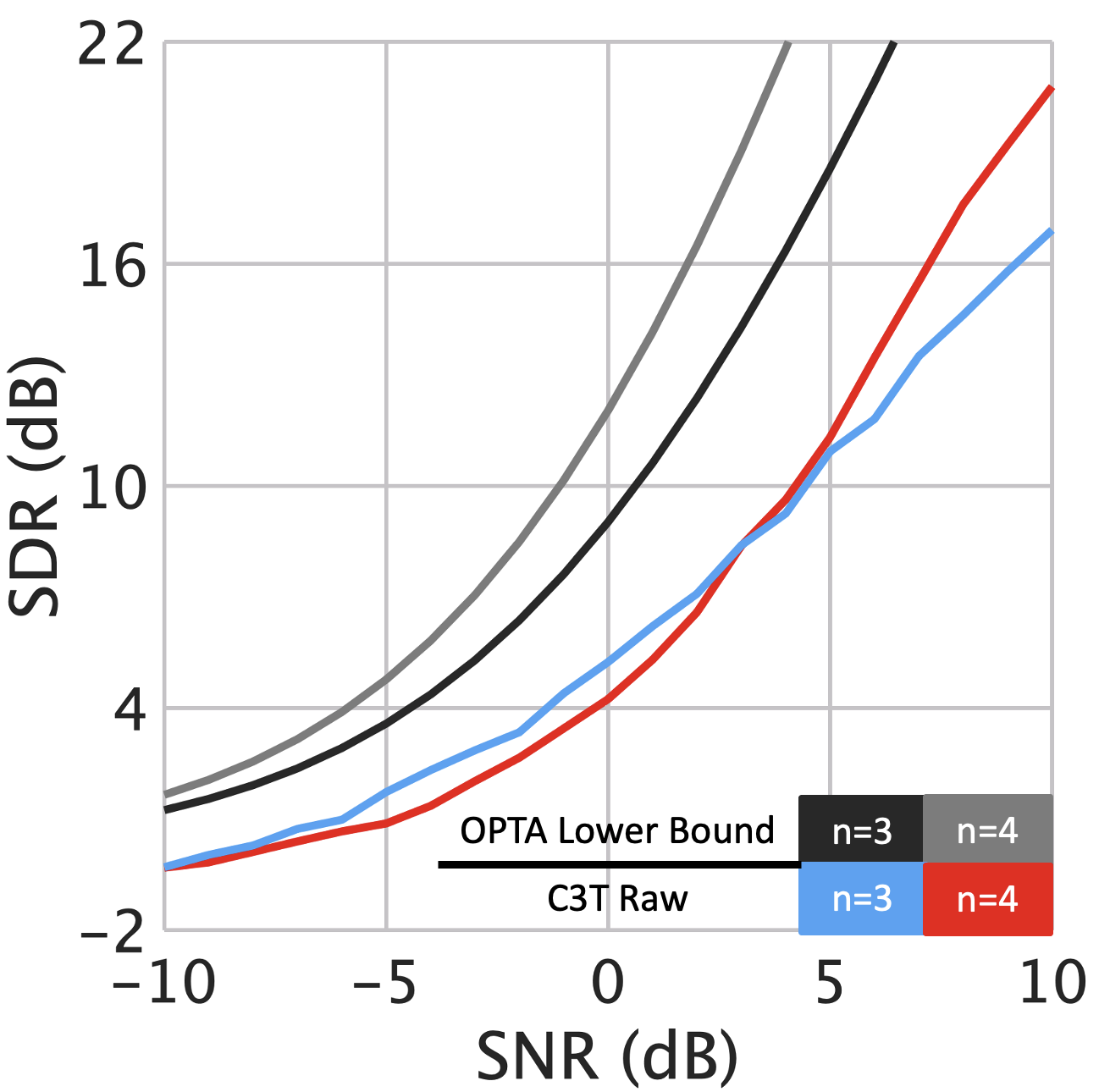}
    \caption{Performance of the proposed C3T codes for $n = 3$ and $4$ compared with the two associated OPTA lower bounds.}
    \vspace{-0.5cm}
    \label{fig:perf_n3}
\end{figure}

In Figs. \ref{fig:perf_n2n4n6n8}, \ref{fig:perf_n20n40n100},  and \ref{fig:perf_n3}, we plot \textrm{SDR} versus \textrm{SNR} for our proposed analog coding scheme. For reference, we also show the OPTA lower bound \eqref{equ:opta_lower_bound}, the MAP decoder \eqref{equ:map}, the MMSE decoder \eqref{equ:mmse}, the repetition code sample mean decoder, and the repetition code linear MMSE (LMMSE) decoder performance. A repetition code \cite{Bossert1999} entails repeating the symbol to be transmitted $n$ times, where $n$ corresponds to the C3T dimension. We derive the LMMSE decoder applied to the repetition code in Appendix \ref{sec:rep_mmse}. We illustrate the performance curves for the three cases of the MLP decoder described in Section~\ref{sec:decoder}: Raw observations, TP and AO. Additionally, we show the performance curves for four combinations of the aforementioned features, i.e., Raw+TP, Raw+AO, TP+AO, and Raw+TP+AO.  

%-----------------------------------------------------------------------------------
%-----------------------------------------------------------------------------------
%-----------------------------------------------------------------------------------
%-----------------------------------------------------------------------------------

As expected, the MLP decoder performs reasonably well in the low \textrm{SNR} region. The feature combinations that involve AO features have a much shorter inference time and perform significantly better than the ones using raw observations, TP, and Raw+TP cases at \textrm{SNRs} below $0$ dB. For example, the $100$-dimensional Raw+TP+AO case produces an \textrm{SDR} performance of $+13$ dB at an \textrm{SNR} level of $\minus5$ dB, which is $11$ dB better than the Raw Observations case.  At high \textrm{SNRs}, the $40$-dimensional Raw, TP and Raw+TP cases achieve exceedingly high \textrm{SDR} levels. The channel \textrm{SNR} at which we experience the elbow in the curves is a consequence of the maximization of the minimum circumradius function and occurs around $5\num{-10}$ dB \textrm{SNR}.

The repetition code with the LMMSE decoder performs the best of all decoders at \textrm{SNR} levels below $0$ dB for $n = 2$, $4$, $6$, and $8$. At extremely low SNRs, it approaches the OPTA. On the other hand, at higher \textrm{SNR}s or higher coding dimensions, the repetition code LMMSE performance is similar to that of the mean estimator. 

For odd BW expansion factors, Fig.~\ref{fig:perf_n3} shows the performance of the MLP decoder for $n=3$ using raw observations. The MLP decoder follows OPTA up to about $0$ dB \textrm{SNR}. Above $+3$ dB \textrm{SNR}, the performance of the $n=4$ case exceeds that of $n=3$, as expected because of the larger bandwidth expansion factor. However, the $n=3$ case performs better below $+3$ dB \textrm{SNR}. This could be attributed to the fact that in the odd-dimensional case, the geometry is a helix and not a flat torus as in the even-dimensional case. In other words, different geometries result in different slopes of the \textrm{SDR} curve.
%-----------------------------------------------------------------------------------
%-----------------------------------------------------------------------------------

The performance of the MAP decoder starts to level out at a certain \textrm{SNR} level depending on the dimension. This is due to the implementation of the decoder, which correlates the received vector with a set of pre-computed vectors that represent a subset of all possible transmitted vectors for a particular dimension and feature set.  Note that the size of this subset may be set large enough to achieve some desired maximum error. The MAP decoder finds the strongest correlation peak and the source estimate is the source value associated with the vector yielding that strongest peak.  When the AWGN of the channel shrinks below a certain point, the same correlation peak is almost always identified and the mean-squared error of the source estimate is not reduced as the AWGN is reduced below that point. We observe that for all BW expansion factors, the MLP decoders outperform the MAP decoder at low \textrm{SNRs}. This is especially true for very large values of the BW expansion factor. For instance, when $n = 100$, using the RAW+TP+AO or RAW+AO features at $−10$ dB \textrm{SNR}, the MLP decoders shows nearly $8$ dB improvement over the MAP decoder. Note that, for uniformly distributed sources, the posterior distribution for the ML decoder is the same as the MAP decoder because the prior distribution is constant for uniform i.i.d. random variables. The MMSE decoder has a similar performance as the MAP decoder. 

\begin{figure*}[t]
\centering
    \includegraphics[width=1.0\textwidth]{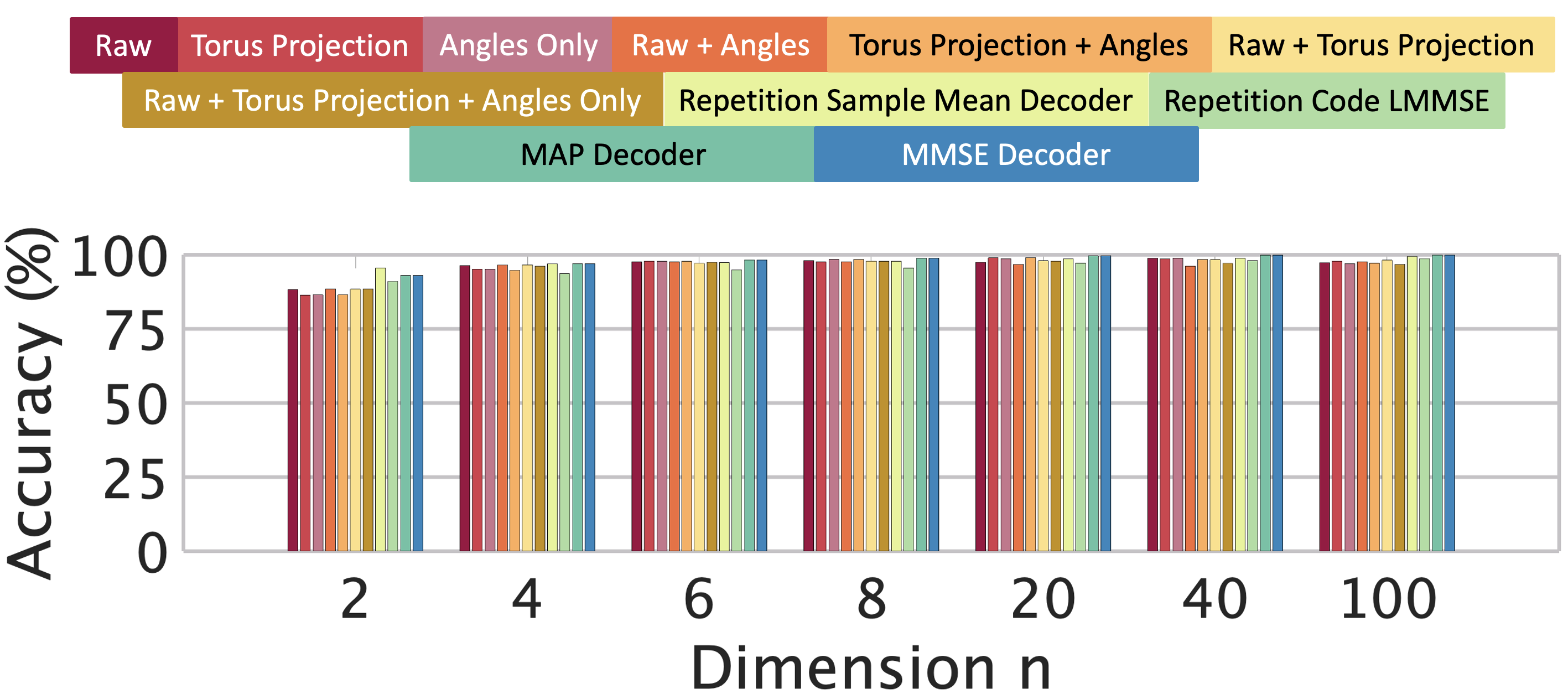}
    %\vspace{-1cm}
    \caption{Accuracy of the MLP decoder for seven combinations of feature sets for $n= 2, 4, 6, 8, 20, 40$, and $100$ compared with the repetition code sample mean decoder, repetition code LMMSE decoder, MAP decoder, and C3T MMSE decoder at $10$ dB \textrm{SNR}.}
    \label{fig:accuracy}
\end{figure*}

Fig. \ref{fig:accuracy} shows the C3T source estimation accuracy at $10$ dB \textrm{SNR} for all dimensions and feature sets. The accuracy for $n=3$ using raw observations is about $94$\% at $10$ dB \textrm{SNR}; this is slightly worse than $n=4$.  The highest MLP decoder accuracy at $10$ dB \textrm{SNR} is achieved using the RAW, TP, or a combination of these two features.  For comparisons with the torus projection method used by other geometrical codes, we consider the mentioned constant curvature codes presented in \cite{VaishampayanCosta2003} (hereafter, V\&C codes, after the initials of the authors). This means that the code is geometrically uniform. Another related work is \cite{Campello2013}, but it uses several constant curvature curves in multiple tori on the sphere and a comparison is not straightforward. It should be interesting to discuss in future works how the approach presented here could be used in this context.  

\begin{figure}[t]
\centering
    \includegraphics[width=1.0\columnwidth]{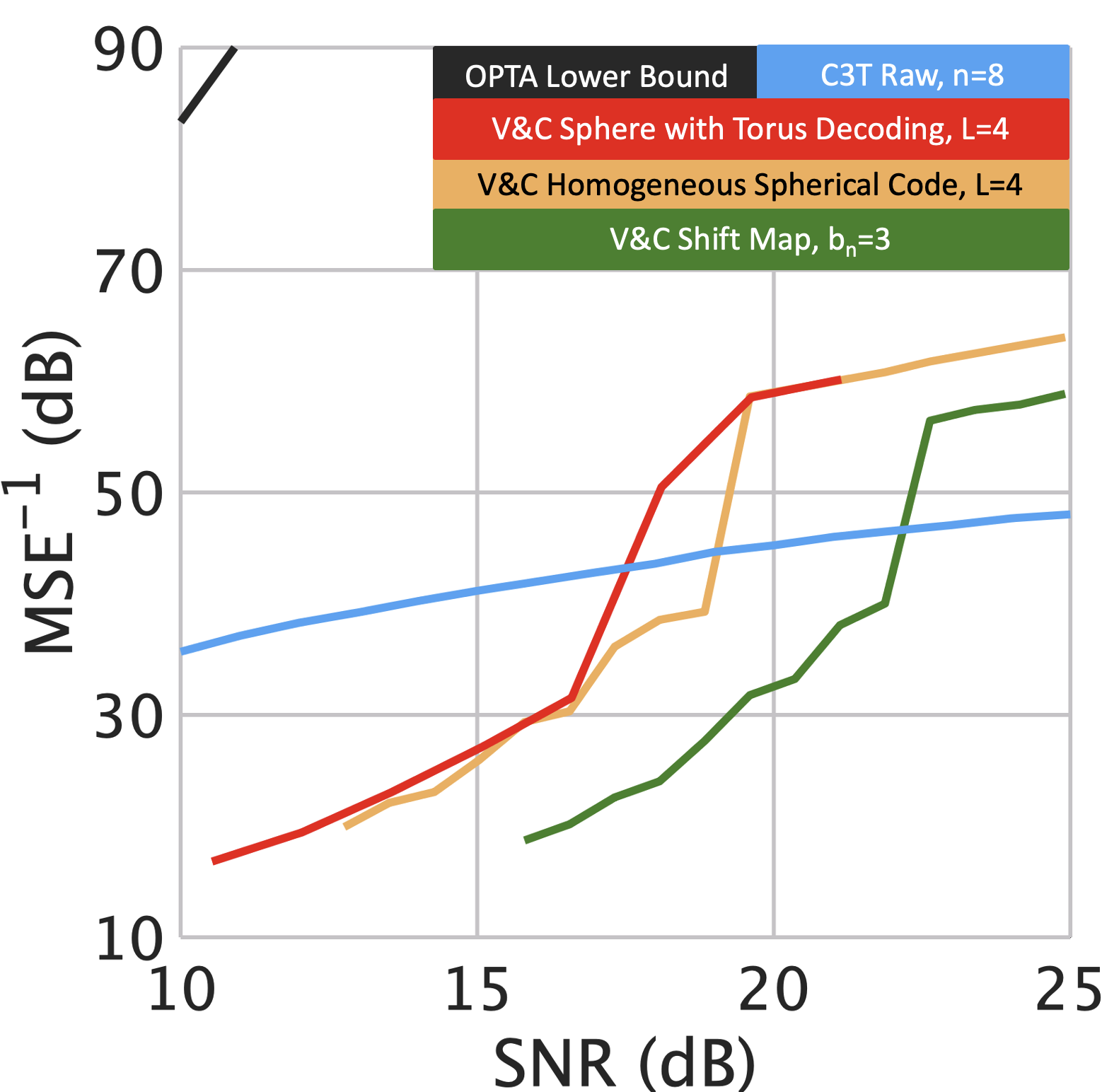}
    %\vspace{-1cm}
    \caption{Reciprocal of MSE for the MLP decoder with $n=8$; the $(1,3,9,27)$ homogeneous spherical code \cite{VaishampayanCosta2003} with conventional and torus decoders; and shift-map codes \cite{VaishampayanCosta2003} for uniformly i.i.d sources.}
    \label{fig:perf_v_and_c}
\end{figure}

Fig.~\ref{fig:perf_v_and_c} compares the MLP decoder ($n=8$) with that of $(1,3,9,27)$-spherical code with the conventional and the torus decoder \cite{VaishampayanCosta2003}. We also plot the performance of shift-map code \cite{VaishampayanCosta2003} for \emph{uniform i.i.d. sources} with interval length $0.5$.  The encoder of the spherical code consists of four $\{\textrm{cos}/\textrm{sin}\}$ pairs. This corresponds to a BW expansion factor of $8$. For $\textrm{SNR} < 17$ dB, the C3T torus projection decoder performs significantly better than both the V\&C torus decoder and shift map. Note that the $\textrm{SDR} = \sigma^2_s / D$ includes the source variance $\sigma^2_s = 1/3$. For performance comparisons without source variance, $4.77$ dB should be added to $\textrm{MSE}^{-1}$.

%-----------------------------------------------------------------------------------
%-----------------------------------------------------------------------------------
%-----------------------------------------------------------------------------------
%-----------------------------------------------------------------------------------
    
In general, the accuracy seems to improve as the number of dimensions increases, however, for the case of $n = 100$ dimensions, the accuracy is lower for the TP features and the RAW+TP combination.  This could be a result of not properly optimizing the learning rate, and/or an insufficient number of epochs. This also explains the generally lower SDR levels for $n = 100$ in Fig. \ref{fig:perf_n20n40n100}. We did an experiment with doubling the hidden MLP layers and this resulted in a modest performance improvement of $1.5$ dB for $n = 100$.

\begin{figure}[t]
\centering
    \includegraphics[width=1.0\columnwidth]{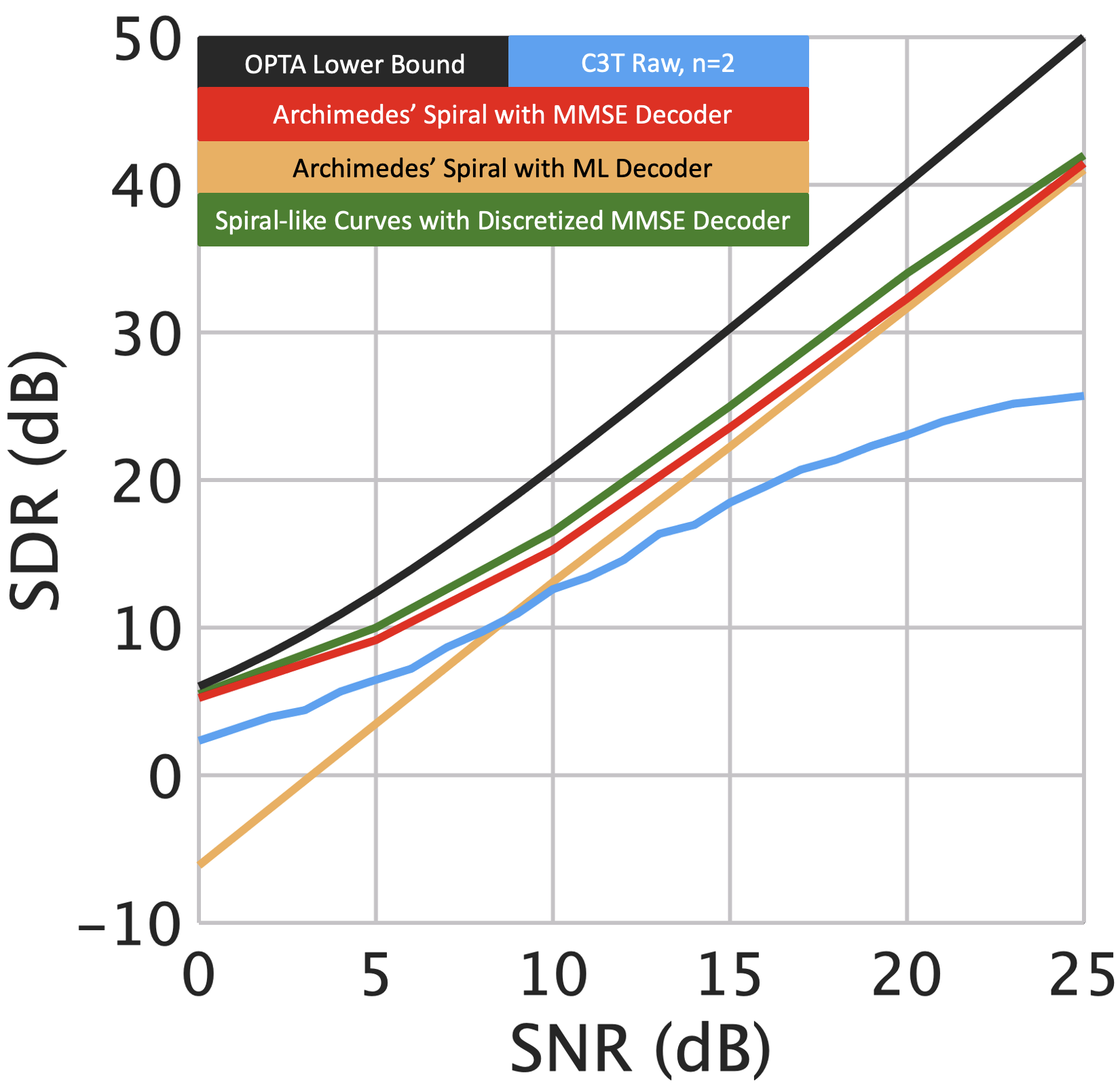}
    %\vspace{-1cm}
    \caption{Performance of the proposed C3T codes for $n=2$ compared with that of the $1$:$2$ Archimedes' spiral mapping with MMSE and ML decoders \cite{Hekland2009} and spiral-like curves with discretized MMSE decoder \cite{Saleh2011}. In each case, \emph{uniformly} i.i.d. sources were used.}
    \label{fig:archimedes}
\end{figure}

Here, it is instructive to compare our proposed C3T codes with the non-linear analog codes such as the $1$:$2$ Archimedes' spiral mapping \cite{Hekland2009}, where uniformly i.i.d. source samples are mapped to a point on the double Archimedes' spiral generating a two-dimensional symbol. Fig.~\ref{fig:archimedes} shows that, for $\textrm{SNR} < 9$ dB, the MLP decoder with $n=2$ performs better than the $1$:$2$ Archimedes' spiral with ML decoder. However, when the MMSE decoder is employed for the $1$:$2$ Archimedes' spiral, its performance is better than the MLP decoder for the entire range of \textrm{SNR}. Fig.~\ref{fig:archimedes} also depicts the performance of spiral-like curves from \cite{Saleh2011} with discretized MMSE decoder. This analog code is based on the double-intertwined Archimedes’ spiral with nearly uniformly spaced spiral arms. 
%-----------------------------------------------------------------------------------
%-----------------------------------------------------------------------------------

\subsection{Comparison with optimal digital codes}
We use the lower bound on the channel code length for digital coding in \eqref{equ:pb} to compute the required number of source symbols needed for digital codes compared to a single source symbol needed for our analog C3T code. Table II shows the message length expansion factor needed by digital codes to match the performance of the C3T code for fixed quantities of BW expansion factor, \textrm{SNR}, and \textrm{SDR}. We force the number of channel uses per analog source sample to be constant between the analog and digital schemes by setting the digital code rate R equal to the number of required quantization bits divided by the number of channel-use $n$ of the C3T code, i.e. $R = (B + 1) /n$. The required number of source samples $N_s$ (message block-length before coding divided by the number of quantization bits) for Block Error Rates (BLER) of $\epsilon = 1e\num{-3}$ and $\epsilon = 1e\num{-6}$ are displayed in the last two columns of the table. Since the required number of source samples for the C3T analog
code in each case is one, these numbers represent the message expansion factor needed for the optimal digital code to match the performance of the C3T code. As an example, looking at the C3T code with $\textrm{SNR} = 3.0$ dB and $\textrm{SDR} = 20$ dB in Table {\ref{tab:block_length}}, the best digital code requires a code rate of $R = 0.72$ to match the (\textrm{SNR},\textrm{SDR}) point that uses six channels per source sample.  The digital code requires queuing up $686$ source samples before applying the rate $0.72$ code to achieve a \textrm{SDR} of $20$ dB at a \textrm{SNR} of $3.0$ dB. This illustrates the unique and powerful performance advantage that analog codes can have in the short block-length regime, which is a strict requirement for URLLC applications.
%-----------------------------------------------------------------------------------
\begin{table}[t]
    \centering
    \caption{\footnotesize{Number of source samples $N_s$ needed to achieve BLER, $\epsilon$, for particular \textrm{SNR} and \textrm{SDR} values.}}
    \begin{tabular}{cccccc}
      \toprule
      %\genfrac #1 #2 {0pt}{2}{a+b}{c+d}
      $n$ & \textrm{SNR} (dB) & \textrm{SDR} (dB) & R & $\genfrac{}{}{0pt}{2}{N_s}{\epsilon = 10^{-3}}$ & $\genfrac{}{}{0pt}{2}{N_s}{\epsilon = 10^{-6}}$ \\
      \midrule
        4 & 0.5 & 12 & 0.748 & 45 & 108     \\
        4 & 3.69 & 18 & 0.997 & 138 & 328   \\
        6 & 3.0 & 20 & 0.720 & 290 & 686    \\
        6 & 4.98 & 25 & 0.859 & 55 & 132    \\
        6 & 8.39 & 30 & 0.997 & 6 & 15      \\
        8 & 3.57 & 24 & 0.623 & 20 & 49     \\
        8 & 5.95 & 30 & 0.748 & 7 & 17      \\  
        20 & 0.36 & 24 & 0.249 & 4 & 11     \\  
        20 & 2.89 & 36 & 0.349 & 2 & 5      \\
        100 & -5.45 & 20 & 0.043 & 2 & 4    \\
      \bottomrule
    \end{tabular}
    \label{tab:block_length}
\end{table}
%-----------------------------------------------------------------------------------

For URLLC applications the requirement to wait for nearly $700$ samples before transmission would be prohibitive. If we look closely at Table \ref{tab:radii_params}, it is easy to spot that the lower-dimensional C3T codes show the strongest advantage over digital codes. Since the gap to OPTA widens for the higher dimensional C3T codes, this observation is not surprising. The high-dimensional C3T codes are necessary to get adequate \textrm{SDRs} at very low \textrm{SNR}. Although the improvement factor in message queue length in those higher dimensional cases is only single digits, the overall computational complexity needed to reach those target \{channel uses per source sample, \textrm{SNR},
\textrm{SDR}\} tuples is less than digital coding encoders and decoders. Furthermore, these comparisons are to the optimal digital channel code and, hence, the actual improvement factors to real practical channel codes are likely much higher.

\subsection{Geometrical vs autoencoder-based encoders}%\label{sec:analysis}

We also verified if an autoencoder framework may successfully learn
a better encoder map than a geometrically-designed counterpart such as
the C3T encoder. The autoencoder enjoys the freedom of not
being restricted to a curve of constant curvature. So, it would seem
in principle to have a good chance. To explore this
concept, we parameterized the encoder and decoder functions by two fully
connected MLP neural networks which are trained
jointly and considered as an autoencoder with a non-trainable
layer in the middle that represents the noisy communications channel.

A deep MLP autoencoder neural network is a nested
nonlinear function approximator. In principle, an ideal function
that accurately captures the encoder, channel, and decoder is
guaranteed to exist by the universality theorem for neural networks
~\cite{Cybenko1994} and by the fact that multilayer feedforward
neural networks are universal function approximators
~\cite{Hornik1989}.

We observed that the autoencoder framework under thousands of
different random weight initializations never discovered an
encoder-decoder network that improved upon the geometrically
structured C3T code in terms of \textrm{SDR} vs \textrm{SNR} performance.  As pointed
out in \cite{Choromanska2015}, recovering the global minimum becomes
harder as the network size increases. Further, the
objective function used by a neural network is analogous to the
Hamiltonian of the spin-glass model under the assumptions of variable independence, redundancy in network parametrization, and uniformity \cite{Choromanska2015}. All of these conditions are met in our simple MLP networks. In \cite{Choromanska2015},  the lowest critical values of the Hamiltonians of these models were shown to form a layered structure and located in a well-defined band lower-bounded by the global minimum.  In our experiments, the encoder portion of our autoencoder network does not
deviate significantly from the weights initialized using the constant curvature curve encoder approximation function.  Furthermore, initializing the weights randomly over many different seeds, failed to approach the performance of the constant curvature curve seeded autoencoder. 

Three autoencoder-based schemes for low-delay analog joint source-channel coding have been explored for Gaussian and Gauss-Markov sources over AWGN channels [24]. The autoencoders were built using recurrent neural networks and these deep-learning methods yield SDR close to or better than the Archimedes spirals [13], especially at low SNRs.

\section{Summary}
\label{sec:summ}
We proposed a novel framework for analog error correction of i. i. d. continuous uniform sources based on densely packing the hypertubes inside of hyperspheres. The center line of the designed hypertube is a constant curvature curve corresponding to the image of the encoder map. In our numerical experiments using an MLP-based decoder, various combinations of C3T features with AO features were shown to perform considerably well at lower \textrm{SNR} levels. The low-dimension C3T codes are a superior alternative in low-latency applications because they produce acceptable \textrm{SDRs} without the need for very long code block-lengths to maintain a specified block error rate. 

In general, we have shown that C3T codes are both reliable and operate with extremely low-latency. Using high-dimensional codes (in the range of $n=20$ to 40) and at high \textrm{SNR} levels (around $10$  dB), we achieve very high source estimation accuracy with the \textrm{SDR} approaching $35$ dB by decoding the raw channel output or TP thereof. The accuracy improves as the number of dimensions increases. The low latency is inherent in these types of codes because of the unity block-length. This property makes such codes attractive for URLLC applications.

\appendices
\section{Proof of Theorem~\ref{theorem:alpha_given_y}}
\label{sec:theorem_alpha_given_y_proof}

\begin{IEEEproof}
  Using the joint density of $(y_{2i-1},y_{2i})$, we write the joint density of $(\eta_i,\beta_i)$ as
  \begin{equation}
    p_{\beta_i,\eta_i}(\beta_i,\eta_i) = \beta_i p_{y_{2i-1},y_{2i}}(\beta_i\cos(\eta_i),\beta_i\sin(\eta_i)) .
  \end{equation}
  From here, the marginal for $\eta_i$ is 
  \begin{align*}
        p_{\eta_i} & = \int_{\beta_i=0}^{\infty} p_{\beta_i,\eta_i}(\beta_i,\eta_i)  d\beta_i \\ 
    & \propto \int_{\beta_i=0}^{\infty} \beta_i\exp\left\{-\frac{1}{2\sigma_w^2}(\beta_i\cos(\eta_i)-r_i\cos(\omega_i\alpha))^2\right\} \\ 
    &  \exp\left\{-\frac{1}{2\sigma_w^2}(\beta_i\sin(\eta_i)-r_i\sin(\omega_i\alpha))^2\right\} d\beta_i \\ 
    & \propto \int_{\beta_i=0}^{\infty} \beta_i\exp\left\{ -\frac{\beta_i^2}{2\sigma_w^2}\right\}
    \exp\left\{\frac{1}{\sigma_w^2}(\beta_ir_i\cos(\eta_i)\cos(\omega_i\alpha)\right. \\ 
    & \left.+ \beta_ir_i\sin(\eta_i)\sin(\omega_i\alpha))\right\}  d\beta_i \\ 
    & \propto \int_{\beta_i=0}^{\infty} \beta_i\exp\left\{ -\frac{\beta_i^2}{2\sigma_w^2}\right\}
    \exp\left\{\frac{\beta_ir_i}{\sigma_w^2}\cos(\eta_i-\omega_i\alpha) \right\} d\beta_i  \\
    & \propto \exp\left(\frac{r_i^2\cos^2(\eta_i-\omega_i\alpha)}{2\sigma_w^2}\right) \\ 
    & \int_{\beta_i=0}^{\infty} \beta_i \exp\left\{ -\frac{1}{2\sigma_w^2} (\beta_i-r_i\cos(\eta_i-\omega_i\alpha))^2\right\}  d\beta_i  \\ 
    & \propto \exp\left(\frac{r_i^2\cos^2(\eta_i-\omega_i\alpha)}{2\sigma_w^2}\right) \\ 
    & \left[\sigma_w^2 \exp\left\{ -\frac{1}{2\sigma_w^2} r_i^2\cos^2(\eta_i-\omega_i\alpha)\right\} \right.\\
        &  + \sigma_w\frac{\pi}{2}r_i\cos(\eta_i-\omega_i\alpha) \\
        & \left.\left(1+\mbox{erf}\left(\frac{r_i\cos(\eta_i-\omega_i\alpha)}{\sqrt{2\sigma_w^2}}\right)\right) \right] \\
    & \propto 1 + \sqrt{\pi}q_i(\alpha)\exp(q_i^2(\alpha))\mbox{erfc}(-q_i(\alpha)), \;\;\;
  \end{align*}
  where $q_i(\alpha)=\frac{r_i\cos(\eta_i-\omega_i\alpha)}{\sqrt{2\sigma_w^2}}$.
\end{IEEEproof}

\section{LMMSE for Repetition Code}
\label{sec:rep_mmse}
Consider an $n$-dimensional repetition code of a source signal, $s$, over an AWGN channel. The Bayesian linear model for the received signal $\textbf{y}$ is
    \begin{align*}
        \mathbf{y} = s\mathbf{1}_n + \mathbf{w}, 
    \end{align*}
    where $\mathbf{w} \sim \mathcal{N}(0,\sigma^2_w \mathbf{I})$, $s \sim \mathcal{U}(-a,+a)$, $\mathrm{E}[s] = 0$, and $\mathrm{E}[\textbf{y}] = \mathbf{0}_n$.  Given the prior pdf, $p(s) = \frac{1}{2a} $, define %the variance of $s$, the cross-covariance $\textbf{C}_{sy}$, and the covariance $\textbf{C}_{yy}$ are respectively,
    \begin{align*}
        &\sigma^2_s = \mathrm{E}[s^2] = \int_{-a}^{+a} s^2p(s) ds = \int_{-a}^{+a} s^2\frac{1}{2a} ds = \frac{s^3}{6a} \bigg|^{+a}_{-a} = \frac{a^2}{3},\\
        &\textbf{C}_{sy} = \mathrm{E}[s\mathbf{y}^T] = \mathrm{E}[s(s\mathbf{1}_n + \mathbf{w})^T] = \sigma^2_s\mathbf{1}^T_n \\
        &\textbf{C}_{yy} = \mathrm{E}[\mathbf{yy}^T ] = \mathrm{E}[(s\mathbf{1}_n + \mathbf{w})(s\mathbf{1}_n + \mathbf{w})^T] = \sigma^2_s\mathbf{1}_n\mathbf{1}^T_n + \sigma^2_w \mathbf{I}.
    \end{align*}
    The MMSE of $s$, also known as the \emph{minimum Bayesian mean square error estimator}, is the expectation of the posterior pdf $p(s|\mathbf{y})$:
    \begin{align*}
        \hat{s} & = \mathrm{E}[s|\mathbf{y}] \\
        & = \int_{-\infty}^{+\infty} sp(s|\mathbf{y})ds \; .
    \end{align*}
    The posterior pdf is typically found using the joint pdf $p(\mathbf{y},s)$ as
    \begin{align*}
        p(s|\mathbf{y}) = \frac{p(\mathbf{y},s)}{p(\mathbf{y})} \; .
    \end{align*}
    Note that here $p(\mathbf{y},s)$ is not Gaussian. Hence, we restrict the estimator to be linear resulting in the following LMMSE of $s$ \cite{Kay2013}:
    \begin{align*}
        \hat{s} & = \mathrm{E}[s] + \textbf{C}_{sy}\textbf{C}^{-1}_{yy}(\mathbf{y}-\mathrm{E}[\mathbf{y}]) \\
        & = \sigma^2_s\mathbf{1}^T_n(\sigma^2_s\mathbf{1}_n\mathbf{1}^T_n + \sigma^2_w \mathbf{I})^{-1}\mathbf{y} \\
        & = \frac{na^2}{na^2+3\sigma^2_w}\overline{y} \;,
    \end{align*}
    which is essentially a practical Wiener filter. The LMMSE requires that the observed data $y$ is correlated with the parameters to be estimated $s$. In case of a large dimension $n$ or for high \textrm{SNR}, the estimator $\hat{s}$ tends towards the sample mean $ \overline{y}$. A realization of this estimator would use the median in lieu of the sample mean in order to filter out potential outliers.

%\clearpage
%\balance
\bibliographystyle{IEEEtran}
\bibliography{main}

\end{document}